\documentclass[a4paper,11pt]{article}
\pdfoutput=1 
\usepackage{jcappub}
\usepackage{amsmath}
\usepackage{graphicx}

\usepackage{geometry}
\usepackage{xspace}
\usepackage{pdflscape}

\makeatletter
\gdef\@fpheader{}
\g@addto@macro\bfseries{\boldmath}
\makeatother

\newcommand{\ie}{{i.e.~}}

\newcommand{\eg}{\textsl{e.g.~}}

\newcommand{\order}[1]{\mathcal{O}\!\left(#1\right)}

\DeclareMathOperator{\erfc}{erfc}

\newcommand{\dd}{\mathrm{d}}
\newcommand{\ee}{e}

\newcommand{\sss}[1]{{\scriptscriptstyle{#1}}}

\newcommand{\uPl}{\mathrm{Pl}}

\newcommand{\umin}{\mathrm{min}}

\newcommand{\uend}{\mathrm{end}}

\newcommand{\ueff}{\mathrm{eff}}

\newcommand{\usssPl}{\sss{\uPl}}

\newcommand{\Mp}{M_\usssPl}

\newcommand{\efolds}{$e$-folds~}
\newcommand{\efold}{$e$-fold}

\newcommand{\beq}{\begin{equation}}
\newcommand{\eeq}{\end{equation}}
\newcommand{\bea}{\begin{equation}\begin{aligned}}
\newcommand{\eea}{\end{aligned}\end{equation}}

\newlength{\wsingfig}
\newlength{\wdblefig}
\newlength{\wquadfig}
\newlength{\wtriplefig}
\newcommand{\Eq}[1]{Eq.~(\ref{#1})}
\newcommand{\Eqs}[1]{Eqs.~(\ref{#1})}
\newcommand{\Fig}[1]{Fig.~{\ref{#1}}}

\newcommand{\Ref}[1]{Ref.~{\cite{#1}}}
\newcommand{\Refs}[1]{Refs.~{\cite{#1}}}
\newcommand{\Sec}[1]{Sec.~\ref{#1}}
\newcommand{\Secs}[1]{Secs.~\ref{#1}}
\newcommand{\App}[1]{Appendix~\ref{#1}}
\newcommand{\Apps}[1]{Appendixes~\ref{#1}}

\newcommand{\DKL}{D_{\mathrm{KL}}}
\newcommand{\DJS}{D_{\mathrm{JS}}}

\subheader{}

\title{The stochastic spectator}

\author[a]{Robert J. Hardwick,}
\author[a]{Vincent Vennin,}
\author[b]{Christian T. Byrnes,}
\author[b]{Jes\'us Torrado,}
\author[a]{and David Wands}

\affiliation[a]{Institute of Cosmology \& Gravitation, University of Portsmouth, Dennis Sciama Building, Burnaby Road, Portsmouth, PO1 3FX, United Kingdom}
\affiliation[b]{Department of Physics \& Astronomy, University of Sussex, Brighton BN1 9QH, United Kingdom}

\emailAdd{robert.hardwick@port.ac.uk}
\emailAdd{vincent.vennin@port.ac.uk}
\emailAdd{c.byrnes@sussex.ac.uk}
\emailAdd{jesus.torrado@sussex.ac.uk}
\emailAdd{david.wands@port.ac.uk}

\date{today}

\begin{document}
\sloppy

\abstract{We study the stochastic distribution of spectator fields predicted in different slow-roll inflation backgrounds. Spectator fields have a negligible energy density during inflation but may play an important dynamical role later, even giving rise to primordial density perturbations within our observational horizon today. During de-Sitter expansion there is an equilibrium solution for the spectator field which is often used to estimate the stochastic distribution during slow-roll inflation. However slow roll only requires that the Hubble rate varies slowly compared to the Hubble time, while the time taken for the stochastic distribution to evolve to the de-Sitter equilibrium solution can be much longer than a Hubble time. We study both chaotic (monomial) and plateau inflaton potentials, with quadratic, quartic and axionic spectator fields. We give an adiabaticity condition for the spectator field distribution to relax to the de-Sitter equilibrium, and find that the de-Sitter approximation is never a reliable estimate for the typical distribution at the end of inflation for a quadratic spectator during monomial inflation. The existence of an adiabatic regime at early times can erase the dependence on initial conditions of the final distribution of field values. In these cases, spectator fields acquire sub-Planckian expectation values. Otherwise spectator fields may acquire much larger field displacements than suggested by the de-Sitter equilibrium solution. We quantify the information about initial conditions that can be obtained from the final field distribution. Our results may have important consequences for the viability of spectator models for the origin of structure, such as the simplest curvaton models.}



\maketitle

\section{Introduction}
\label{sec:Introduction}
\label{sec:SpectatorExamples}
Inflation~\cite{Starobinsky:1980te, Sato:1980yn, Guth:1980zm, Linde:1981mu, Albrecht:1982wi, Linde:1983gd} is a phase of accelerated expansion at very high energy in the primordial Universe. During this epoch, vacuum quantum fluctuations of the gravitational and matter fields were amplified to large-scale cosmological perturbations~\cite{Starobinsky:1979ty, Mukhanov:1981xt, Hawking:1982cz,  Starobinsky:1982ee, Guth:1982ec, Bardeen:1983qw}, that later seeded the cosmic microwave background anisotropies and the large-scale structure of our Universe. At present, the full set of observations can be accounted for in a minimal setup, where inflation is driven by a single scalar inflaton field with canonical kinetic term, minimally coupled to gravity, and evolving in a flat potential in the slow-roll regime~\cite{Martin:2013tda, Martin:2013nzq}. From a theoretical point of view, however, inflation takes place in a regime that is far beyond the reach of accelerators, and the physical details of how the inflaton is connected with the standard model of particle physics and its extensions are still unclear. In particular, most physical setups that have been proposed to embed inflation contain extra scalar fields. This is notably the case in string theory models where many extra light moduli fields may be present~\cite{Turok:1987pg, Damour:1995pd, Kachru:2003sx, Kofman:2003nx, Krause:2007jk, Baumann:2014nda}.

Even if such fields are purely spectators during inflation (\ie contribute a negligible amount to the total energy density of the Universe), they can still play an important dynamical role afterwards. The details of their post-inflationary contribution typically depend on the field displacement they acquire during inflation. In this context, if inflation provides initial conditions for cosmological perturbations, it should also be seen as a mechanism that generates a distribution of initial field displacements for light degrees of freedom. In this paper, we investigate what possibilities this second channel offers to probe the physics of inflation. In practice, we study how the field value acquired by light scalar spectator fields at the end of inflation depends on the  inflaton field potential, on the spectator field potential and on the initial distribution of spectator field values.

As an illustration of post-inflationary physical processes for which the field value acquired by spectator fields during inflation plays an important role, we may consider the curvaton scenario \cite{Linde:1996gt, Enqvist:2001zp, Lyth:2001nq, Moroi:2001ct, Bartolo:2002vf}. In this model, the curvaton field $\sigma$ is a light spectator field during inflation that can dominate the energy budget of the Universe afterwards. Its density perturbation is given by $\delta \rho_\sigma/\rho_\sigma\sim \delta\sigma/\sigma$, where $\rho_\sigma$ denotes the energy density contained in $\sigma$, and the effect of this perturbation on the total density perturbation of the Universe is reduced by the relative energy density of the curvaton field to the total energy density. The curvaton field, like every light scalar field, is perturbed at Hubble radius exit by an amount $\delta\sigma\sim H_*\lesssim 10^{-6} \Mp$, where $H_*$ is the Hubble parameter evaluated at the time of Hubble radius crossing during inflation and $\Mp$ is the reduced Planck mass. If the curvaton perturbations produce the entire observed primordial density perturbation with amplitude $10^{-5}$, the average field value in our Hubble patch, $\sigma$, is of order $\sigma\sim 10^5 H_* $. An important question is therefore whether such a field value can naturally be given to the curvaton during inflation. In the limit of low energy scale inflation in particular, this implies that $\sigma \ll \Mp$. The requirement for a very sub-Planckian spectator field value in models where an initially isocurvature field perturbation is later converted into the observed adiabatic curvature perturbation is common but not completely generic, and may be intuitively understood by realising that if the spectator field fluctuations are negligible compared to the background value (\ie $\delta\sigma< 10^{-5} \sigma$), then it is difficult to make the primordial density perturbation have a significant dependence on $\delta\sigma$ if the background value is not very sub-Planckian. This is discussed in the conclusions of \Ref{Byrnes:2008zz}, which shows that it typically also applies to scenarios such as modulated reheating \cite{Dvali:2003em,Kofman:2003nx}. The dark energy model proposed in \Ref{Ringeval:2010hf} also requires sub-Planckian spectator fields during inflation, and the new results we derive on the field value distribution of a spectator field with a quartic potential may have implications for the stability of the Higgs vacuum during inflation as well, see \eg \Refs{Espinosa:2007qp,Herranen:2014cua,Kearney:2015vba}.

This naturally raises the question of whether having a sub-Planckian spectator field value represents a fine tuning of the initial conditions or not. Provided that inflation lasts long enough, we address this question here by calculating the stochastically generated distribution of spectator field values. We will show cases in which sub-Planckian field values are natural, and others in which super-Planckian field values are preferred.

If the spectator field value is driven to become significantly super-Planckian, it can drive a second period of inflation, which may have observable effects even if the inflaton field perturbations dominate, because the observable scales exit the Hubble radius at a different time during the first period of inflation, when the inflaton is traversing a different part of the potential~\cite{Vennin:2015vfa, Vennin:2015egh}. In some cases, we will show that the spectator field value may naturally become so large that it drives more than 60 \efolds of inflation. In this case we would not observe the initial period of inflation at all,  but its existence remains important for generating the initial conditions for the second, observable period of inflation.

If no isocurvature perturbations persist after reheating, the linear perturbations from the inflaton and spectator field are likely to be observationally degenerate. Non-linear perturbations, especially the coupling between primordial long- and short-wavelength perturbations, help to break this degeneracy. We will not study non Gaussianity in this paper, but highlight that the results calculated here help to motivate a prior distribution for the initial spectator field value, which is a crucial ingredient of model comparison between single- and multiple-field models of inflation \cite{Hardwick:2015tma,Vennin:2015vfa, Vennin:2015egh, dePutter:2016trg}.

The paper is organised as follows. In \Sec{sec:StochasticInflation}, the stochastic inflation formalism is introduced. This framework allows one to compute the field value acquired by a quantum scalar field on large scales during inflation, in the form of a classical probability distribution for the field values. In de-Sitter space-times, an equilibrium solution exists for this distribution that is often used as an estimate for more generic inflationary backgrounds as long as they are in the slow-roll regime, hence close to de Sitter. In \Sec{sec:validity-stat}, we explain why this ``adiabatic'' approximation is in fact not valid in general, and why the de-Sitter equilibrium cannot even be used as a proxy in most relevant cases (for a discussion on de-Sitter equilibrium for non-minimally coupled spectators, see \Refs{GarciaBellido:1993wn, GarciaBellido:1994vz, GarciaBellido:1995br}). This is why in \Secs{sec:quad_spec}, \ref{sec:quart_spec} and~\ref{sec:axion_spec}, we study the stochastic dynamics of spectator fields with quadratic, quartic and axionic (\ie cosine) potentials respectively. Each case is investigated with two classes of inflationary potentials, namely plateau and monomial. In \Sec{sec:sigmaendpriors}, the amount of information about the state of the spectator field at the onset of inflation that can be extracted from its field value at the end of inflation is quantified and discussed. Finally, in \Sec{sec:conclusions}, we summarise our main results and draw a few conclusions. Various technical results are derived in \Apps{sec:quadmoments} and~\ref{sec:nonlin_drift}. 

%
\subsection{Stochastic inflation}
\label{sec:StochasticInflation}
Let us now see how the field value acquired by quantum scalar fields during inflation can be calculated in practice. During inflation, scalar field perturbations are placed in squeezed states, which undergo quantum-to-classical transitions~\cite{Polarski:1995jg, Lesgourgues:1996jc, Kiefer:2008ku, Burgess:2014eoa, Martin:2015qta, Boddy:2016zkn} in the sense that on super-Hubble scales, the non-commutative parts of the fields become small compared to their anti-commutative parts. This gives rise to the stochastic inflation formalism~\cite{Starobinsky:1982ee, Starobinsky:1986fx, Nambu:1987ef, Nambu:1988je, Kandrup:1988sc, Nakao:1988yi, Nambu:1989uf, Mollerach:1990zf, Linde:1993xx, Starobinsky:1994bd, Finelli:2008zg, Finelli:2010sh}, consisting of an effective theory for the long-wavelength parts of the quantum fields, which are ``coarse grained'' at a fixed physical scale larger than the Hubble radius during the whole inflationary period. In this framework, the short wavelength fluctuations behave as a classical noise acting on the dynamics of the super-Hubble scales as they cross the coarse-graining scale.  The coarse-grained fields can thus be described by a stochastic classical theory, following Langevin equations
\bea
\label{eq:Langevin}
\frac{\dd\sigma}{\dd N}=-\frac{V_{,\sigma}(\sigma)}{3H^2}+\frac{H}{2\pi}\xi\, .
\eea
In this expression, $\sigma$ denotes a coarse-grained field with potential $V(\sigma)$ (a subscript ``${,\sigma}$'' corresponds to partial derivation with respect to $\sigma$). The Hubble parameter  $H$ is defined as $H\equiv \dot{a}/a$, where $a$ is the scale factor and a dot denotes differentiation with respect to cosmic time. The time variable $N\equiv \ln(a)$ has been used but the choice of the time variable is irrelevant for test fields~\cite{Finelli:2008zg, Finelli:2010sh, Finelli:2011gd, Vennin:2015hra}. Finally, $\xi$ is a Gaussian white noise with vanishing mean and unit variance such that $\langle \xi(N) \rangle = 0$ and $\langle \xi(N_1)\xi(N_2)\rangle = \delta(N_1-N_2)$, where $\langle\cdot\rangle$ denotes ensemble average. The Langevin equation~(\ref{eq:Langevin}) is valid at leading order in slow roll and perturbation theory~\cite{Seery:2007wf, Seery:2010kh}, hence for a light test field with $V_{,\sigma\sigma}\ll H$.  In the It\^o interpretation, it gives rise to a Fokker-Planck equation for the probability density $P(\sigma,N)$ of the coarse-grained field $\sigma$ at time $N$~\cite{Starobinsky:1986fx, Vilenkin:1999kd}
\bea
\label{eq:FP}
\frac{\partial P(\sigma,N)}{\partial N} = \frac{\partial}{\partial\sigma}\left[\frac{V_{,\sigma}(\sigma)}{3H^2} P(\sigma,N)\right]+\frac{H^2}{8\pi^2}\frac{\partial^2}{\partial\sigma^2}\left[ P(\sigma,N)\right]\,.
\eea
This equation can be written as $\partial P/\partial N = -J_{,\sigma}$, where $J\equiv -V_{,\sigma} P/(3H^2)-H^2 P{,_\sigma}/(8\pi^2) $ is the probability current.

When $H$ is constant, a stationary solution $P_\mathrm{stat}$ to \Eq{eq:FP} can be found as follows. Since  $P_\mathrm{stat}$ does not depend on time, the probability current does not depend on $\sigma$ (or on time either). Therefore, if $J$ vanishes at the boundaries of the field domain, it vanishes everywhere. This yields a first-order differential equation for $P_\mathrm{stat}(\sigma)$ that can be solved and one obtains
\bea
P_\mathrm{stat}(\sigma)\propto \exp\left[-\frac{8\pi^2V(\sigma )}{3H^4} \right]\, ,
\label{eq:Pstat}
\eea
where the overall integration constant is fixed by requiring that the distribution is normalised, $\int P(\sigma)\dd\sigma = 1$. In the following, the solution~(\ref{eq:Pstat}) will be referred to as the ``de-Sitter equilibrium''. For instance, if the spectator field has a quadratic potential $V(\sigma)=m^2\sigma^2/2$, the de-Sitter equilibrium is a Gaussian with standard deviation $\sqrt{\langle \sigma^2 \rangle} \sim H^2/m$. In this case, it will be shown in \Sec{sec:plateau_quad_spec} that this equilibrium solution is in fact an attractor of \Eq{eq:Pstat}, that is reached over a time scale $N_\mathrm{relax}\sim H^2/m^2$. Therefore, provided inflation lasts more than  $N_\mathrm{relax}$ $e$-folds, the typical field displacement is of order $H^2/m$ at the end of inflation in this case~\cite{Enqvist:2012xn}.
\subsection{Limitations of the adiabatic approximation}
\label{sec:validity-stat}
In the absence of more general results prior to this paper, the de-Sitter results derived in \Sec{sec:StochasticInflation} have been commonly used and/or assumed to still apply to more realistic slow-roll backgrounds, see \eg \Refs{Enqvist:2012xn,Enqvist:2013kaa,Herranen:2014cua,Hardwick:2015tma,Vennin:2015egh,Figueroa:2016dsc}. The reason is that $H$ varies slowly during slow-roll inflation, which thus does not deviate much from de Sitter. This is why in practice, \Eq{eq:Pstat} is often used to estimate the field value acquired by spectator fields during inflation. However, one can already see why this ``adiabatic'' approximation, which assumes that one can simply replace $H$ by $H(N)$ in \Eq{eq:Pstat} and track the local equilibrium at every time, is not always valid. Indeed,   the time scale over which $H$ varies by a substantial amount in slow-roll inflation is given by $N_H=1/\epsilon_1$, where $\epsilon_1\equiv -\dot{H}/H^2$ is the first slow-roll parameter. During inflation, $\epsilon_1\ll 1$, so that $N_H\gg 1$. However, in order to see whether a spectator field tracks the de-Sitter equilibrium, one should not compare $N_H$ to $1$, but to $N_\mathrm{relax}$, the number of \efolds required by the spectator field to relax towards the equilibrium. In other words, only if the adiabatic condition
\bea
\label{eq:adiabatic:condition}
N_\mathrm{relax} \ll N_H=\frac{1}{\epsilon_1}
\eea
holds can $H$ be considered as a constant over the time required by the spectator field to relax to the equilibrium, and only in this case can the stationary distribution~(\ref{eq:Pstat}) be used.

If the inflaton potential is of the plateau type and asymptotes to a constant as the field value asymptotes to infinity, one typically has~\cite{Roest:2013fha, Martin:2016iqo} $\epsilon_1\simeq \order{1}/(N_\uend-N)^2$ in the limit where $N_\uend-N\gg 1$, where $N_\uend$ denotes the number of \efolds at the end of inflation where $\epsilon_1\simeq 1$. This leads to
\bea
H\simeq H_\mathrm{plateau} \exp\left[{\frac{\order{1}}{N-N_\uend-1}}\right]\, ,
\eea
where $H_\mathrm{plateau}$ is the asymptotic value of $H$ at large-field value, hence $N_H\simeq \order{1} (N_\uend-N)^2$, meaning that $H$ cannot change by more than a factor of order one throughout the entire inflationary phase. For instance, if one considers the Starobinsky potential~\cite{Starobinsky:1980te} $V(\phi)\propto (1-\ee^{-\sqrt{2/3}\phi/\Mp})^2$, one finds $\epsilon_1\simeq 3/[4(N_\uend-N)^2]$ and $H_\uend/H_\mathrm{plateau}\simeq 0.53$. In this case, the de-Sitter equilibrium~(\ref{eq:Pstat}), $\langle V(\sigma) \rangle \sim H^4$, only changes by a relatively small fraction and therefore provides a useful estimate for the order of magnitude of spectator field displacements at the end of inflation [using either $H=H_\mathrm{plateau}$ or $H=H_\uend$ in \Eq{eq:Pstat}]. Note that the same can be true for hilltop potentials where $H$ also asymptotes a constant in the infinite past.

In the context of single-field inflation however, plateau potentials are known to provide a good fit to the data only in the last $\sim \! 50$ \efolds of inflation. The shape of the inflaton potential is not constrained beyond this range and is typically expected to receive corrections when the field varies by more than the Planck scale. In multiple-field inflation, observations allow the inflaton potential to be of the large-field type all the way down to the end of inflation~\cite{Vennin:2015egh}. Therefore we also consider monomial inflaton potentials $V(\phi)\propto \phi^p$ with $p>0$. In these models, one has
\bea
\label{eq:Hubble}
H(N)=H_\uend\left[1+\frac{4}{p}\left(N_\uend-N\right)\right]^{\frac{p}{4}}\, .
\eea
If $p>1$, this corresponds to convex inflaton potentials (meaning $V''>0$), while this describes concave inflaton potentials ($V''<0$) for $p<1$, and the de-Sitter case is recovered in the limit $p\rightarrow 0$. From \Eq{eq:Hubble}, one has $\epsilon_1=(H_\uend/H)^{4/p}$, so that $N_H=(H/H_\uend)^{4/p}$. If the spectator field has a quadratic potential for instance, as mentioned above, it will be shown in \Sec{sec:plateau_quad_spec} that $N_\mathrm{relax}\sim H^2/m^2$. In this case, the adiabatic condition~(\ref{eq:adiabatic:condition}) reads $(H/H_\uend)^{2/p-1}\gg H_\uend/m$. If $p\geq 2$, one can see that this can never be realised since $H_\uend>m$ and $H>H_\uend$. If $p <2$, the adiabatic condition is satisfied when $H$ is sufficiently large, that is to say at early enough times when $N_\uend-N>p[(H_\uend/m)^{4/(2-p)}-1]/4$. If $m/H_\uend \sim 0.01$ for instance, this number of \efolds is larger than $\sim 400$ as soon as $p>0.1$ (and larger than $\sim 10^7$ for $p>1$), which means that even in this case, the adiabatic regime lies far away from the observable last $50$ \efolds of inflation. One concludes that in most cases, the de-Sitter equilibrium solution does not provide a reliable estimate of the field value acquired by spectator fields during inflation. In the following, we therefore study the dynamics of such fields beyond the adiabatic approximation.
\section{Quadratic spectator}
\label{sec:quad_spec}
In this section, we consider a quadratic spectator field, for which
\bea
V(\sigma)=\frac{m^2}{2}\sigma^2\, .
\eea
In this case, the Langevin equation~(\ref{eq:Langevin}) is linear, which allows one to solve it analytically. In \App{sec:quadmoments}, we explain how to calculate the first two statistical moments of the spectator field $\sigma$. The first moment is given by
\bea
\label{eq:meansigma}
\langle\sigma\left(N\right)\rangle = \langle\sigma\left(N_0\right)\rangle\exp\left[-\frac{m^2}{3}\int_{N_0}^N\frac{\dd{N}^\prime}{H^2({N}^\prime)}\right]\,,
\eea
which corresponds to the classical solution of \Eq{eq:Langevin} in the absence of quantum diffusion, and where we have set $\langle \sigma \rangle = \langle \sigma(N_0) \rangle $ at the initial time $N_0$. For the second moment, one obtains
\bea
\label{eq:meansigmasquare:final}
\left\langle \sigma^2(N)\right\rangle =& \left\langle \sigma^2(N_0)\right\rangle \exp\left[-\frac{2m^2}{3}\int_{N_0}^{N}\frac{\dd N^{\prime}}{H^2(N^{\prime})}\right]
\\ &
+ \int_{N_0}^N \dd N^\prime \frac{H^2(N^\prime)}{4\pi^2} \exp\left[\frac{2m^2}{3}\int_N^{N^\prime}\frac{\dd N^{\prime\prime}}{H^2(N^{\prime\prime})}\right]\, .
\eea
In this expression, the structure of the first term in the right-hand side is similar to the first moment~(\ref{eq:meansigma}) while the second term is due to quantum diffusion, so that the variance of the distribution $\langle \sigma^2 \rangle-\langle\sigma\rangle^2$ is given by the same formula as the second moment [\ie one can replace $\langle \sigma^2\rangle$ by $\langle \sigma^2 \rangle-\langle\sigma\rangle^2$ in \Eq{eq:meansigmasquare:final} and the formula is still valid].

One can also show that the Fokker-Planck equation~(\ref{eq:FP}) admits Gaussian solutions,
\bea
\label{eq:ansatz:Gaussian}
P\left(\sigma,N\right)=\frac{1}{\sqrt{2\pi\left\langle \sigma^2(N) \right\rangle}}\exp\left\lbrace-\frac{\left[\sigma-\left\langle\sigma(N)\right\rangle\right]^2}{2\left\langle \sigma^2 \right\rangle}\right\rbrace\, ,
\eea
where $\langle \sigma(N) \rangle$ and  $\langle \sigma^2(N) \rangle$ are given by \Eqs{eq:meansigma} and~(\ref{eq:meansigmasquare:final}) respectively. However, let us stress that \Eqs{eq:meansigma} and~(\ref{eq:meansigmasquare:final}) are valid for any (\ie not only Gaussian) probability distributions.
\subsection{Plateau inflation}
\label{sec:plateau_quad_spec}
As explained in \Sec{sec:validity-stat}, if the inflaton potential is of the plateau type, $H$ can be approximated by a constant. In this case, the mean coarse-grained field~(\ref{eq:meansigma}) is given by
\bea
\left\langle \sigma(N) \right\rangle =
\left\langle \sigma(N_0) \right\rangle
\exp\left[-\frac{m^2}{3H^2}\left(N-N_0\right)\right]
\, .
\eea
It follows the classical trajectory as already pointed out below \Eq{eq:meansigma}, and becomes small when $N-N_0\gg H^2/m^2$. For the second moment, \Eq{eq:meansigmasquare:final} gives rise to
\bea
\left\langle \sigma^2(N) \right\rangle = \left[\left\langle \sigma^2(N_0) \right\rangle - \frac{3H^4}{8\pi^2m^2}\right]
\exp\left[-\frac{2m^2}{3H^2}\left(N-N_0\right)\right] + \frac{3H^4}{8\pi^2m^2}\, .
\eea
When $N-N_0\gg H^2/m^2$, it approaches the constant value $\langle \sigma^2 \rangle = 3H^4/(8\pi^2 m^2)$. One can check that this asymptotic value corresponds to the de-Sitter equilibrium in \Eq{eq:Pstat}. Moreover, one can see that the typical relaxation time that is required to reach the attractor is given by
\bea
\label{eq:quadratic:Nrelax}
N_\mathrm{relax}=\frac{H^2}{m^2}\, ,
\eea
which corresponds to the value reported in \Sec{sec:StochasticInflation}.
\subsection{Monomial inflation}
\label{sec:monomial_quad_spec}
If the inflaton potential is monomial and of the form $V(\phi)\propto \phi^p$, the Hubble factor is given by \Eq{eq:Hubble}. Substituting this expression for $H(N)$ into \Eq{eq:meansigma}, one obtains (for $p\neq 2$)
\bea
\label{eq:meansigma:quadratic:monomial}
\left\langle \sigma(H)\right\rangle = \left\langle \sigma(H_0)\right\rangle \exp\left\lbrace\frac{\mu}{2} \left[\left(\frac{H}{H_\uend}\right)^{\frac{4}{p}-2}-\left(\frac{H_0}{H_\uend}\right)^{\frac{4}{p}-2}\right]\right\rbrace\, ,
\eea
where $H_0$ is the value of $H$ at an initial time $N_0$, and we have defined
\bea
\mu\equiv \frac{m^2}{3H_\uend^2}\frac{p}{2-p}\, .
\eea
In \Eq{eq:meansigma:quadratic:monomial}, time is parametrised by $H$ instead of $N$ for convenience but the two are directly related through \Eq{eq:Hubble}. For the second moment (or for the variance), by substituting \Eq{eq:Hubble} into \Eq{eq:meansigmasquare:final}, one obtains
\bea
&\left\langle \sigma^2(H)\right\rangle = \left\langle \sigma^2(H_0)\right\rangle \exp\left\lbrace \mu \left[\left(\frac{H}{H_\uend}\right)^{\frac{4}{p}-2}-\left(\frac{H_0}{H_\uend}\right)^{\frac{4}{p}-2}\right]\right\rbrace
 \\ & \quad\quad
 +\frac{p H_\uend^2 \mu^{\frac{p+2}{p-2}}}{8\pi^2(p-2)}
\ee^{\mu \left(\frac{H}{H_\uend}\right)^{\frac{4}{p}-2}}
\left\lbrace \Gamma \left[ \frac{2+p}{2-p}, \mu\left(\frac{H_0}{H_\uend}\right)^{\frac{4}{p}-2}\right] -\Gamma \left[ \frac{2+p}{2-p}, \mu\left(\frac{H}{H_\uend}\right)^{\frac{4}{p}-2}\right] \right\rbrace
\label{eq:meansigma2:quadratic:generic}
\eea
where $\Gamma$ denotes the incomplete Gamma function. One can note that both \Eqs{eq:meansigma:quadratic:monomial} and~(\ref{eq:meansigma2:quadratic:generic}) can be expressed as functions $\mu (H/H_\uend)^{4/p-2}$ only, which is directly proportional to the ratio $N_H/N_\mathrm{relax}$. As noted in \Sec{sec:validity-stat}, for $p\geq 2$ this ratio is always small, while for $p<2$, it is large unless $H$ is sufficiently large. The two cases $p\geq2$ and $p<2$ must therefore be treated distinctly.
\subsubsection{Case where $p \geq 2$}
\label{sec:quadratic:pgt2}
If $p>2$, one has  $N_H\ll N_\mathrm{relax}$ and the quantity $\mu (H/H_\uend)^{4/p-2}$ in \Eqs{eq:meansigma:quadratic:monomial} and~(\ref{eq:meansigma2:quadratic:generic}) is always much smaller than one. This implies that the argument of the exponential in \Eq{eq:meansigma:quadratic:monomial} can be neglected, and $\langle \sigma(H) \rangle\simeq \langle \sigma(H_0) \rangle$ stays constant. Therefore, the distribution remains centred at the initial value. Note that the case $p=2$ is singular and gives rise to
\bea
\label{eq:meansigma:quadratic:peq2}
\left\langle \sigma(H)\right\rangle = \left\langle \sigma(H_0) \right\rangle \left(\frac{H}{H_0}\right)^{\frac{m^2}{3H_\uend^2}}\, ,
\eea
which also yields $\langle \sigma(H) \rangle\simeq \langle \sigma(H_0) \rangle$ unless $H_0/H_\uend\gg \exp(3H_\uend^2/m^2)$.

For the second moment, the second arguments of the incomplete Gamma functions in \Eq{eq:meansigma2:quadratic:generic} are always much smaller than one and in this limit, one finds
\bea
\label{eq:meansigma2:quadratic:generic:pGT2}
\left\langle \sigma^2(H)\right\rangle \simeq  \left\langle \sigma^2(H_0)\right\rangle+\frac{H_\uend^2 p}{8\pi^2(p+2)}\left[\left(\frac{H_0}{H_\uend} \right)^{2+\frac{4}{p}}-\left(\frac{H}{H_\uend} \right)^{2+\frac{4}{p}}\right]\, .
\eea
In this expression, one can see that $\langle\sigma^2\rangle$ can only increase as time proceeds, in a way that does not depend on the mass (as long as it is sub-Hubble). The result is therefore the same as if one set the mass to zero, and corresponds to a free diffusion process. This is consistent with the fact that $\langle\sigma\rangle$ stays constant in this case. If $p=2$, \Eq{eq:meansigma2:quadratic:generic} is singular and one has
\bea
\label{eq:meansigma2:quadratic:peq2}
\left\langle \sigma^2(H)\right\rangle = & \left\langle \sigma^2(H_0)\right\rangle  \left(\frac{H}{H_0}\right)^{\frac{2m^2}{3H_\uend^2}}
+\frac{H_\uend^2}{8\pi^2\left(2-\frac{m^2}{3H_\uend^2}\right)}\left(\frac{H}{H_\uend}\right)^{\frac{2m^2}{3H_\uend^2}}
 \\ &
\times\left[\left(\frac{H_0}{H_\uend}\right)^{4-\frac{2m^2}{3H_\uend^2}}-\left(\frac{H}{H_\uend}\right)^{4-\frac{2m^2}{3H_\uend^2}}\right]
\, .
\eea
In this case, it was also shown in \Sec{sec:validity-stat} that $N_H\ll N_\mathrm{relax}$ so there is no adiabatic regime either. Unless $H_0/H_\uend\gg \exp(3H_\uend^2/m^2)$, in the limit $m\ll H_\uend$, \Eq{eq:meansigma2:quadratic:peq2} coincides with \Eq{eq:meansigma2:quadratic:generic:pGT2} evaluated at $p=2$ so in practice the latter formula can be used for all values of $p\geq 2$.

An important feature of \Eq{eq:meansigma2:quadratic:generic:pGT2} is that it strongly depends on the initial conditions $\langle \sigma(H_0)\rangle$ and $H_0$. This is because there is no adiabatic regime in this case and hence no attractor that would erase initial conditions. As a consequence, the typical spectator field displacement at the end of inflation cannot be determined without specifying initial conditions.

One should also note that the present analysis relies on the assumption that the inflaton is not experiencing large stochastic diffusion, which allows us to use \Eq{eq:Hubble}. This is in fact the case if $H\ll H_\mathrm{eternal}$, where
\bea
\label{eq:Hei:def}
H_\mathrm{eternal} \equiv H_\uend \left( \frac{\Mp}{H_\uend}2\pi \sqrt{2} \right)^{\frac{p}{2+p}}
\eea
is the scale above which a regime of so-called ``eternal inflation'' takes place.\footnote
{More precisely, $H_\mathrm{eternal}$ is defined~\cite{Winitzki:2008zz} as the scale above which, over the typical time scale of an $e$-fold, the mean quantum diffusion received by the inflaton field, $H/(2\pi)$, is larger than the classical drift, $\sqrt{2\epsilon_1}\Mp$. Since $\epsilon_1=(H_\uend/H)^{4/p}$ in monomial inflation~(\ref{eq:Hubble}), this condition gives rise to $H>H_\mathrm{eternal}$ where $H_\mathrm{eternal}$ is given by \Eq{eq:Hei:def}.}
For this reason, $H_\mathrm{eternal}$ is the largest value one can use for $H_0$ in order for the calculation to be valid. Setting $H_0=H_\mathrm{eternal}$, and substituting \Eq{eq:Hei:def} into \Eq{eq:meansigma2:quadratic:generic:pGT2}, one obtains at the end of inflation
\bea
\label{eq:quadratic:sigmaend:pgt2}
\left\langle \sigma^2_\uend \right\rangle \simeq \left\langle \sigma^2_\mathrm{eternal} \right\rangle + \frac{p}{p+2}\Mp^2 \, .
\eea
This expression is displayed in the left panel of \Fig{fig:quadratic:summary}. It means that the field value of the spectator field is at least of the order of the Planck mass at the end of inflation. If one assumes the de-Sitter equilibrium distribution~(\ref{eq:Pstat}) at the end of eternal inflation for instance, $\langle \sigma^2_\mathrm{eternal} \rangle = 3H_\mathrm{eternal}^4/(8\pi^2 m^2)$, even much larger field displacements are obtained at the end of inflation.
\subsubsection{Case where $p < 2$}
\label{sec:quadratic:plt2}
\begin{figure}[t]
\begin{center}
\includegraphics[width=0.479\textwidth]{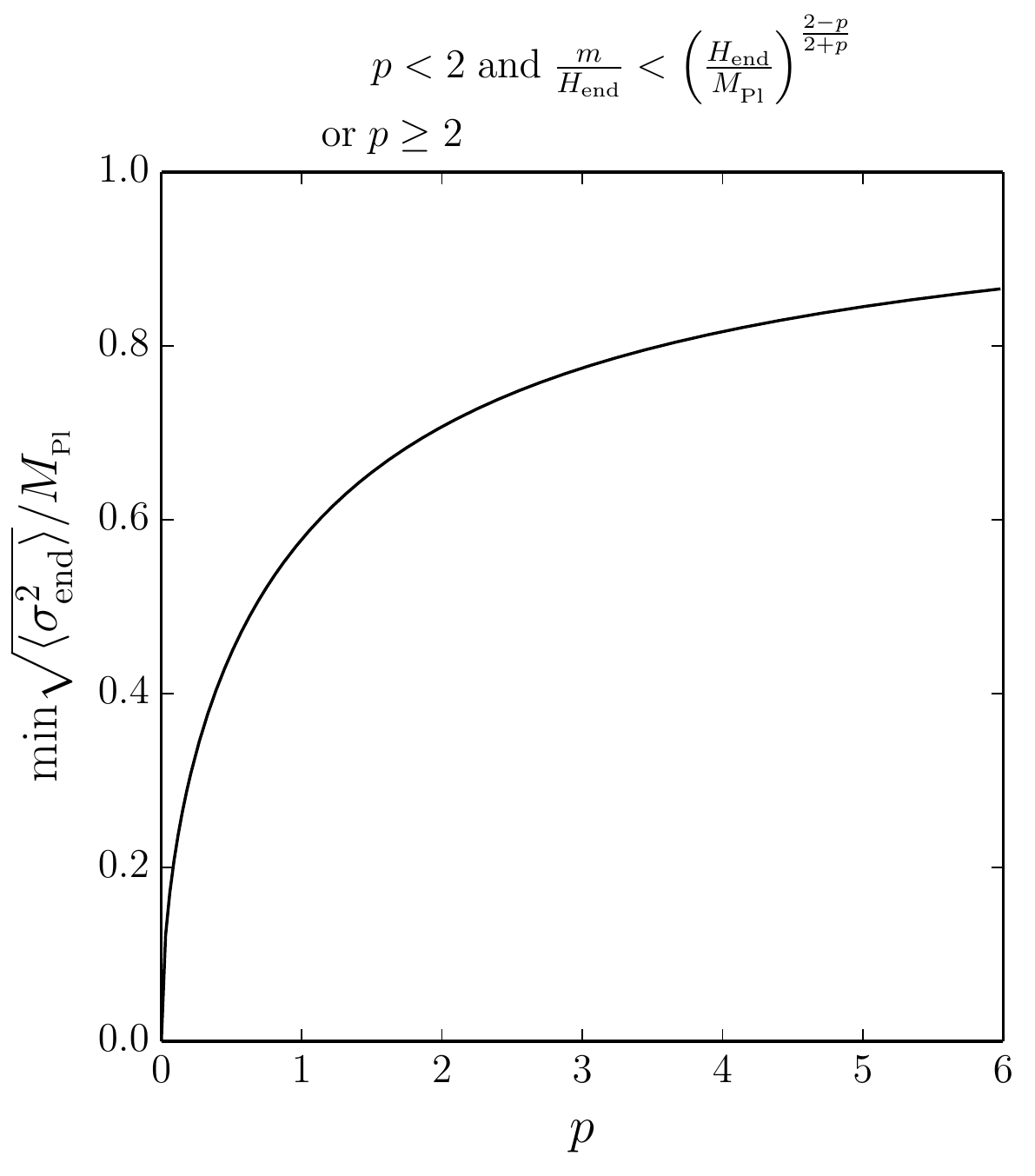}
\includegraphics[width=0.506\textwidth]{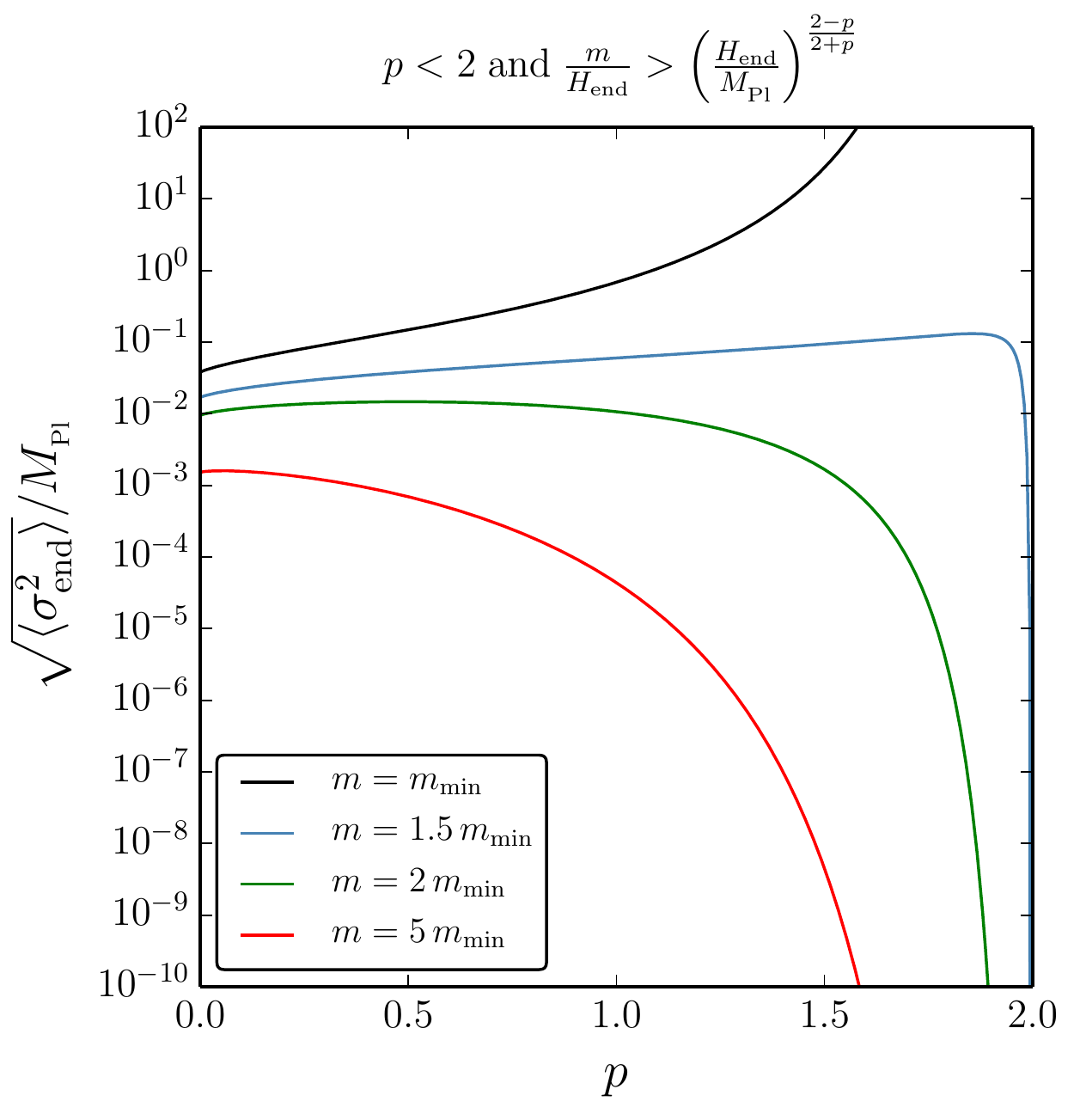}
\caption{The typical field displacement $\sqrt{\langle\sigma_\uend^2 \rangle}$ acquired by a test field $\sigma$ with quadratic potential $V(\sigma)=m^2\sigma^2/2$ at the end of an inflationary phase driven by an inflaton potential $V(\phi)\propto\phi^p$. In the left panel, the cases $p\geq 2$ and $p<2$ with $m/H_\uend<(H_\uend/\Mp)^{(2-p)/(2+p)}$ are displayed, where the minimum value of $\langle\sigma_\uend^2 \rangle$ is given if one initially sets $\langle\sigma^2 \rangle=0$ at the time $H=H_\mathrm{eternal}$ when stochastic corrections to the inflaton dynamics stop being large. This corresponds to \Eq{eq:quadratic:sigmaend:pgt2} and shows that spectator fields are typically at least close to super-Planckian at the end of inflation in these cases. In the right panel, the case $p<2$ with $m/H_\uend>(H_\uend/\Mp)^{(2-p)/(2+p)}$ is displayed, where there is an early adiabatic regime that allows the dependence on initial condition to be erased. The typical field displacement is given by \Eq{eq:quadratic:sigmaend:plt2}, which is expressed as a function of $p$ and $m/m_\umin$ in \Eq{eq:meansigma2:quadratic:plt2:adiabatic:HadiabHeternal}, where $m_\umin$ is the lower bound on $m$ associated to the condition $m/H_\uend>(H_\uend/\Mp)^{(2-p)/(2+p)}$. One can check that, as soon as  $m \gtrsim 1.5\, m_\umin$,  $\sqrt{\langle\sigma_\uend^2 \rangle}$  is always sub-Planckian in this case.
\label{fig:quadratic:summary}}
\end{center}
\end{figure}
If $p<2$, whether the ratio $N_H/N_\mathrm{relax}$ is small or large depends on the value of $H$. More precisely, if $H \gg H_\mathrm{adiab}$, where
\bea
\label{eq:Hadiab:def}
H_\mathrm{adiab} \equiv H_\uend \left(\frac{H_\uend}{m}\right)^{\frac{p}{2-p}}\, ,
\eea
one is in the adiabatic regime and $N_H\gg N_\mathrm{relax}$. As soon as $H$ drops below $H_\mathrm{adiab}$ however, one leaves the adiabatic regime. In order to set initial conditions during the adiabatic regime, it should apply after the eternal inflationary phase during which our calculation does not apply, which implies that $H_\mathrm{adiab}<H_\mathrm{eternal}$. Making use of \Eqs{eq:Hei:def} and~(\ref{eq:Hadiab:def}), this condition gives rise to
\bea
\label{eq:quadratic:adiabstart:condition}
\frac{m}{H_\uend}>\left(\frac{H_\uend}{\Mp}\right)^{\frac{2-p}{2+p}}\, .
\eea
Let us distinguish the two cases where this relation is and is not satisfied.
\paragraph{Starting out in the adiabatic regime}\mbox{} \\
If \Eq{eq:quadratic:adiabstart:condition} is satisfied, one can set initial conditions for the spectator field $\sigma$ in the adiabatic regime while being outside the eternal inflationary phase, that is to say one can take $H_\mathrm{adiab}<H_0<H_\mathrm{eternal}$. From \Eq{eq:meansigma:quadratic:monomial}, this implies that $\langle \sigma_\uend \rangle \ll \langle \sigma_0\rangle$ and the distribution becomes centred around smaller field values as time proceeds. Regarding the width of the distribution, two regimes of interest need to be considered.

At early time, \ie when $H\gg H_\mathrm{adiab}$, the incomplete Gamma functions in \Eq{eq:meansigma2:quadratic:generic} can be expanded in the large second argument limit and one obtains
\bea
\label{eq:meansigma2:quadratic:plt2:adiabatic}
\langle \sigma^2(H) \rangle \simeq \left[\langle \sigma^2(H_0) \rangle - \frac{3H_0^4}{8\pi^2 m^2}\right]\exp\left\lbrace{\mu\left[\left(\frac{H}{H_\uend}\right)^{\frac{4}{p}-2}-\left(\frac{H_0}{H_\uend}\right)^{\frac{4}{p}-2}\right]}\right\rbrace
+ \frac{3H^4}{8\pi^2m^2}\, .
\eea
In this expression, one can see that as soon as $H$ decreases from $H_0$, the first term is exponentially suppressed and one obtains $\langle \sigma^2 \rangle \simeq 3H^4/(8\pi m^2)$, which corresponds to the de-Sitter equilibrium formula\footnote
{More precisely, in a de-Sitter universe where $H$ is constant and equal to the instantaneous value $H(N)$ for a given $N$ in the case at hand, the asymptotic value reached by $\langle\sigma^2\rangle$ at late time is the same as the instantaneous value  $\langle\sigma^2(N)\rangle$ obtained from \Eq{eq:meansigma2:quadratic:plt2:adiabatic}. In this sense, the time evolution of $H$ can be neglected and this corresponds, by definition, to an adiabatic regime.}
and confirms that one is in the adiabatic regime. This also shows that the de-Sitter equilibrium is an attractor of the stochastic dynamics in this case, and that it is reached within a number of \efolds $\sim H_0^2/m^2$, which exactly corresponds to $N_\mathrm{relax}$ given in \Eq{eq:quadratic:Nrelax} when $H=H_0$.

At later times, \ie when $H\ll H_\mathrm{adiab}$, one leaves the adiabatic regime and while the first incomplete Gamma function in \Eq{eq:meansigma2:quadratic:generic} can still be expanded in the large second argument limit, the second one must be expanded in the small second argument limit and this gives rise to
\bea
\label{eq:quadratic:sigmaend:plt2}
\left\langle \sigma^2(H) \right\rangle \simeq  \frac{H_\uend^2}{8\pi^2}\frac{p}{2-p}\Gamma\left(\frac{2+p}{2-p}\right)\left(\frac{3H_\uend^2}{m^2}\frac{2-p}{p}\right)^{\frac{2+p}{2-p}}\, .
\eea
Interestingly, this expression does not depend on $H$, meaning that $\langle \sigma^2 \rangle$ stays constant as soon as one leaves the adiabatic regime (and obviously stops tracking the adiabatic solution). One can also check that in this expression, the limit $p\rightarrow 0$ gives rise to $\langle \sigma^2_\uend\rangle \simeq 3H_\uend^4/(8\pi^2m^2)$, that is to say the de-Sitter equilibrium formula.

An important consequence of this result is that in the case $p<2$ and if $m>m_\umin$, where $m_\umin$ corresponds to the lower bound on $m$ given by \Eq{eq:quadratic:adiabstart:condition}, even if the end of inflation lies far outside the adiabatic regime, the existence of an early adiabatic phase allows initial conditions to be erased. At the end of inflation, the field value of the spectator field only depends on $m$, $H_\uend$ and $p$. This is in contrast with the case $p\geq 2$ where there is no adiabatic regime, even at early time, and initial conditions remain important even at the end of inflation. A second important consequence is that the typical field displacement is always sub-Planckian at the end of inflation in this case. Indeed, substituting the expression given for $m_\umin$ by \Eq{eq:quadratic:adiabstart:condition} into \Eq{eq:quadratic:sigmaend:plt2}, one obtains
\bea
\label{eq:meansigma2:quadratic:plt2:adiabatic:HadiabHeternal}
\frac{\left\langle\sigma^2_\uend\right\rangle}{\Mp^2}=\frac{1}{8\pi^2}\left(\frac{p}{2-p}\right)^{\frac{2p}{p-2}}\Gamma\left(\frac{2+p}{2-p}\right)\left(3 \frac{m_\umin^2}{m^2}\right)^{\frac{2+p}{2-p}}\, .
\eea
This expression is displayed in the right panel of \Fig{fig:quadratic:summary} for a few values of $m/m_\umin$. One can see that as soon as $m\gtrsim 1.5\, m_\umin$, the spectator field is always sub-Planckian at the end of inflation.
\paragraph{Starting out away from the adiabatic regime}\mbox{} \\
If the condition~(\ref{eq:quadratic:adiabstart:condition}) is not satisfied, the adiabatic regime cannot be used to erase initial conditions dependence. If both $H_0$ and $H$ are much smaller than $H_\mathrm{adiab}$, the incomplete Gamma functions in \Eq{eq:meansigma2:quadratic:generic} can be expanded in the small second argument limit and one obtains \Eq{eq:meansigma2:quadratic:generic:pGT2} again. When $H$ becomes small compared to $H_0$, $\langle\sigma^2\rangle$ reaches a constant and the distribution remains frozen until the end of inflation. Letting $H_0=H_\mathrm{eternal}$ as in \Sec{sec:quadratic:pgt2}, this gives rise to \Eq{eq:quadratic:sigmaend:pgt2} and one concludes that, in this case, the spectator field acquires a super-Planckian field value at the end of inflation.
\mbox{}\\
\mbox{}\\
\indent The situation is summarised in the first line of table~\ref{table:summary} in \Sec{sec:conclusions}. If $p<2$ and $m/H_\uend>(H_\uend/\Mp)^{(2-p)/(2+p)}$, quadratic spectator fields acquire sub-Planckian field values at the end of inflation, while if $p\geq 2$ or if $p<2$ with $m/H_\uend<(H_\uend/\Mp)^{(2-p)/(2+p)}$, they are typically super-Planckian.
\subsection{Can a spectator field drive a second phase of inflation?}
\begin{figure}[t]
\begin{center}
\includegraphics[width=0.49\textwidth]{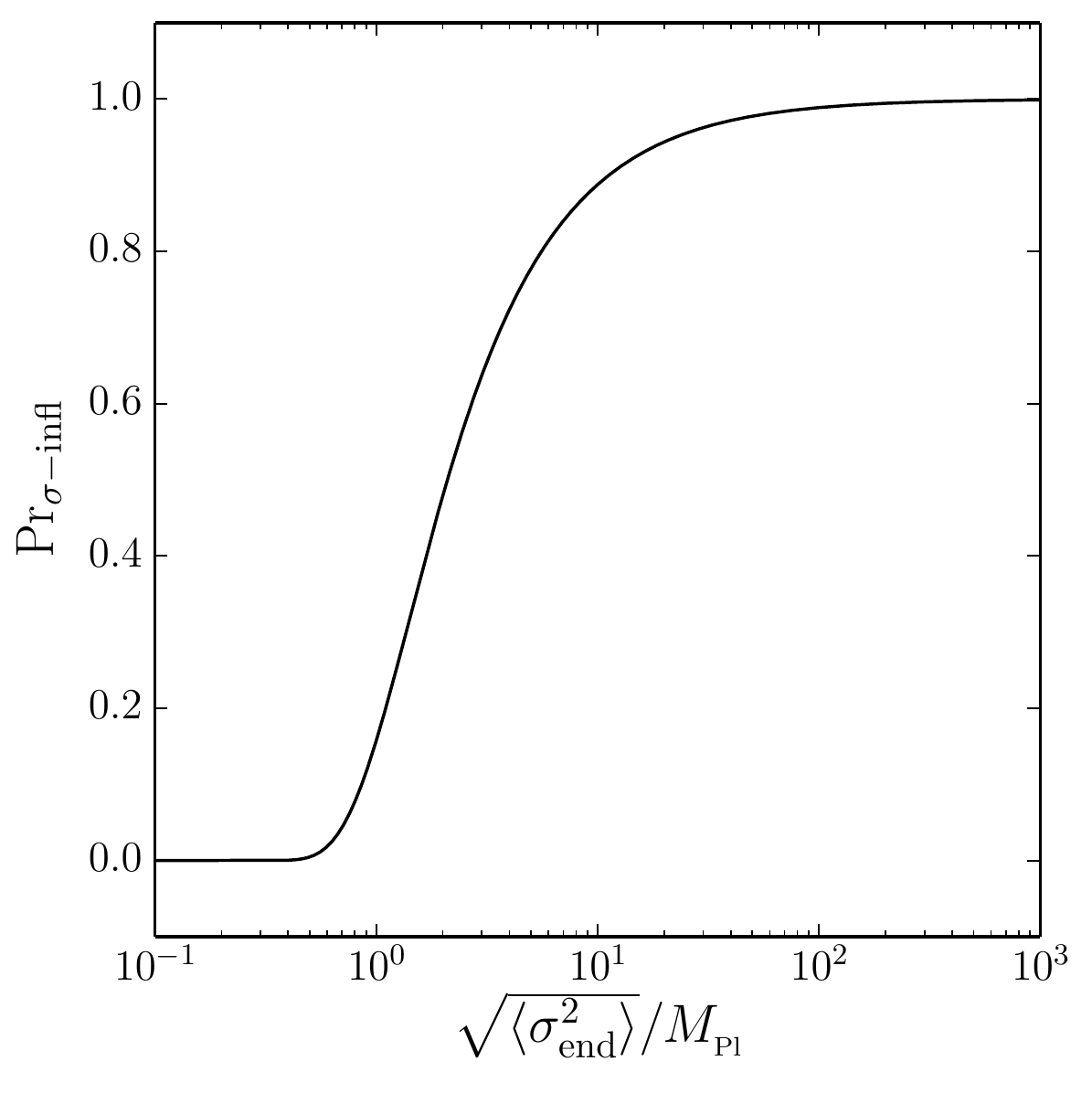}
\includegraphics[width=0.49\textwidth]{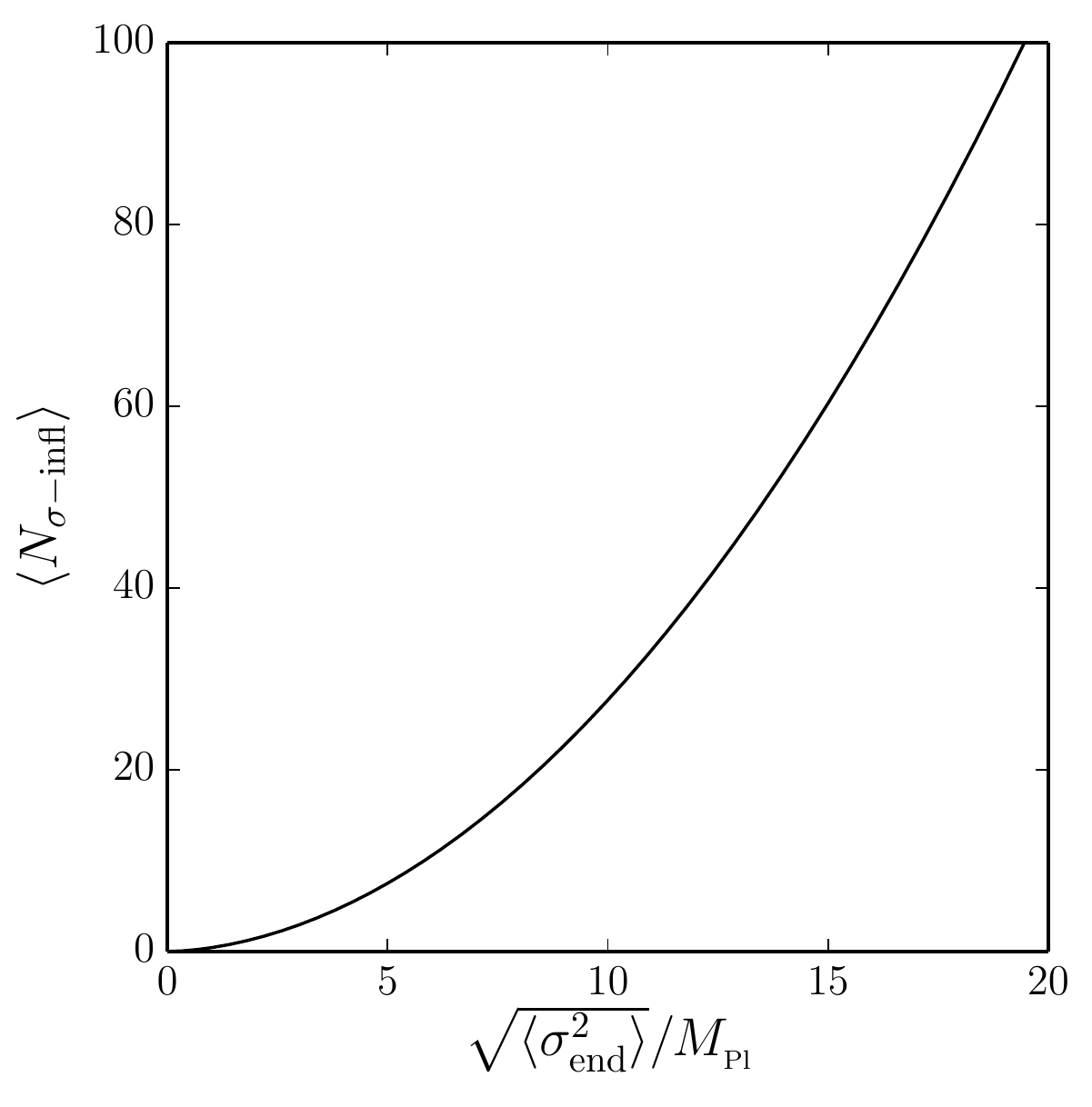}
\caption{A quadratic spectator field $\sigma$ can trigger a second phase of inflation if $\vert \sigma_\uend \vert>\sqrt{2}\Mp$. Assuming a centred Gaussian distribution with variance $\langle\sigma_\uend^2\rangle$, the left panel displays the probability for such a condition to be satisfied, while the mean number of \efolds realised in the second phase of inflation is given in the right panel.
\label{fig:quadratic:SecondInflation}}
\end{center}
\end{figure}
If inflation is driven by a monomial potential $V\propto\phi^p$ with $p\geq2$, in \Sec{sec:monomial_quad_spec} it was shown that quadratic spectator fields typically acquire super-Planckian field values at the end of inflation. This can have important consequences as discussed in \Sec{sec:SpectatorExamples}, amongst which is the ability for the spectator field to drive a second phase of inflation. This can happen if $\vert \sigma_\uend \vert>\sqrt{2}\Mp$, and the probability associated to this condition is given by
\bea
\mathrm{Pr}_{\sigma\text{-}\mathrm{infl}}= \int_{\left\vert\sigma\right\vert > \sqrt{2}\Mp}
P (\sigma,N_\uend)\dd \sigma
= \erfc \left( \frac{\Mp}{\sqrt{\langle \sigma_\uend^2 \rangle}} \right)\, .
\eea
In the second expression, we have assumed that the probability distribution of the spectator field value at the end of inflation is a Gaussian with vanishing mean and variance $\langle\sigma_\uend^2\rangle$, and $\erfc$ denotes the complementary error function. This probability is displayed in the left panel of \Fig{fig:quadratic:SecondInflation}. If a second phase of inflation starts driven by the quadratic potential with initial field value $\sigma_\uend$, then the number of \efolds realised is given by $\sigma_\uend^2/(4\Mp^2)-1/2$. The mean duration of this additional inflationary period can thus be calculated according to
\bea
\langle N_{\sigma\text{-}\mathrm{infl}} \rangle&=   \frac{1}{\mathrm{Pr}\left(\sigma\text{-}\mathrm{infl}\right)}\displaystyle\int_{\left\vert\sigma\right\vert > \sqrt{2}\Mp} \left(\frac{\sigma^2}{4\Mp^2}-\frac{1}{2}\right) P (\sigma,N_\uend)\dd \sigma
\\ &=
\left( \frac{\langle \sigma_\uend^2\rangle}{4\Mp^2} - \frac{1}{2}\right) +
\frac{\sqrt{\langle \sigma_\uend^2 \rangle}}{2\sqrt{\pi}\Mp}
\frac{\exp \left( -\frac{\Mp^2}{\langle \sigma_\uend^2\rangle}\right)}{\erfc \left( \frac{\Mp}{\sqrt{\langle \sigma_\uend^2 \rangle}} \right)} \, ,
\eea
where in the second expression, again, we have assumed that the probability distribution of the spectator field value at the end of inflation is a centred Gaussian. This mean number of \efolds is shown in the right panel of \Fig{fig:quadratic:SecondInflation}. When $\langle \sigma_\uend^2 \rangle$ is super-Planckian, one has a non-negligible probability of a second phase of inflation. For instance, with $\sqrt{\langle \sigma_\uend^2 \rangle }=5\Mp$, one finds $\mathrm{Pr}\left(\sigma\text{-}\mathrm{infl}\right)\simeq 0.77$ and $\langle N_{\sigma\text{-}\mathrm{infl}}\rangle = 7.5$.
\section{Quartic spectator}
\label{sec:quart_spec}
In \Sec{sec:quad_spec}, it was shown that quadratic spectator fields with potential $V(\sigma)=m^2\sigma^2/2$ typically acquire super-Planckian field displacements at the end of inflation if the inflaton potential is of the form $V(\phi)\propto\phi^p$ with $p\geq 2$ at large-field values or with $p<2$ and $m/H_\uend<(H_\uend/\Mp)^{(2-p)/(2+p)}$. In this section, we investigate whether these super-Planckian field values can be tamed by making the spectator field potential steeper at large-field values. In practice, we consider a quartic spectator field,
\bea
\label{eq:quartic:potential}
V(\sigma)=\lambda \sigma^4\, ,
\eea
where $\lambda$ is a dimensionless constant. Contrary to the quadratic case in \Sec{sec:quad_spec}, the Langevin equation~(\ref{eq:Langevin}) is not linear for quartic spectators and cannot be solved analytically. Numerical solutions are therefore presented in this section, where a large number (typically $10^5$ or $10^6$) of realisations of \Eq{eq:Langevin} are generated with a fourth order Runge-Kutta method, over which moments of the spectator field value are calculated at fixed times. These results have been checked with independent numerical solutions of the Fokker-Planck equation~(\ref{eq:FP}).
\subsection{Plateau inflation}
\label{sec:quartic:plateau}
As explained in \Sec{sec:validity-stat}, if the inflaton potential is of the plateau type, $H$ can be approximated by a constant and the spectator field value reaches the de-Sitter equilibrium~(\ref{eq:Pstat}) where the typical field displacement, for the quartic spectator potential~(\ref{eq:quartic:potential}), is given by
\bea
\label{eq:quartic:deSitter}
\left\langle\sigma^2 \right\rangle = \frac{\Gamma\left(\frac{3}{4}\right)}{\Gamma\left(\frac{1}{4}\right)}\sqrt{\frac{3}{2\lambda}} \frac{H^2}{2\pi}\, .
\eea
The relaxation time required to reach this asymptotic value can be assessed as follows. Since the equilibrium~(\ref{eq:Pstat}) is of the form $P(\sigma)\propto e^{-\alpha \sigma^4}$, with $\alpha=8\pi^2\lambda/(3H^4)$, let us assume that the time evolving distribution for $\sigma$ is more generally given by
\bea
\label{eq:quartic:ansatz}
P(\sigma,N)=
\frac{2 \alpha^{1/4}(N)}{\Gamma\left(\frac{1}{4}\right)}\exp\left[-\alpha(N)\sigma^4\right]\, ,
\eea
where $\alpha(N)$ is a free function of time and the prefactor is set so that the distribution remains normalised, and track the stochastic dynamics with this ansatz. By substituting \Eq{eq:quartic:ansatz} into \Eq{eq:FP}, an ordinary differential equation for $\alpha(N)$ is derived in \App{sec:nonlin_drift}, that reads
\bea
\label{eq:quartic:quarticappr:alpha:eom}
\frac{\dd \alpha}{\dd N} &=  \frac{\Gamma \left( \frac{1}{4} \right)}{2\Gamma \left( \frac{3}{4} \right)} \left( \frac{\lambda}{H^2}\alpha^{1/2} - \frac{3H^2}{8\pi^2} \alpha^{3/2}\right) \,.
\eea
If $H$ is a constant, this equation can be solved analytically and the solution is given by \Eq{eq:solution:alpha}. Since \Eq{eq:quartic:ansatz} gives rise to $\langle \sigma^2 \rangle= \alpha^{-1/2} \Gamma(3/4)/\Gamma(1/4)$, one obtains for the second moment
\bea
\label{eq:quartic:plateau:Appr}
\left\langle \sigma^2\left(N\right) \right\rangle = \dfrac{\dfrac{\Gamma\left(\frac{3}{4}\right)}{\Gamma\left(\frac{1}{4}\right)}\sqrt{\frac{3H^4}{8\pi^2\lambda}} }{\tanh\left\lbrace \sqrt{\dfrac{3\lambda}{2}}\dfrac{\Gamma\left(\frac{1}{4}\right)}{8\pi\Gamma\left(\frac{3}{4}\right)}\left(N-N_0\right) +\mathrm{atanh} \left[\sqrt{\dfrac{3H^4 }{8\pi^2\lambda}}\dfrac{\Gamma\left(\frac{3}{4}\right)}{\Gamma\left(\frac{1}{4}\right)\left\langle \sigma^2\left(N_0\right)\right\rangle}\right]  \right\rbrace}\, .
\eea
In the late time limit, one recovers the de-Sitter equilibrium value~(\ref{eq:quartic:deSitter}). Let us stress however that \Eq{eq:quartic:plateau:Appr} is not an exact solution to \Eq{eq:FP} but only provides an approximation under the ansatz~(\ref{eq:quartic:ansatz}). This approximation will be shown to be reasonably accurate in \Sec{sec:quartic:monomial}, but for now, expanding $\tanh(x)\simeq 1-2 \ee^{-2x}$ when $x\gg 1$ at late time, it provides an estimate of the relaxation time as
\bea
\label{eq:quartic:Nrelax}
N_\mathrm{relax} =  \frac{1}{\sqrt{\lambda}}\, .
\eea
It is interesting to notice that this expression is consistent with the numerical exploration of \Ref{Enqvist:2012xn}, see Eq.~(2.12) of this reference.
\subsection{Monomial inflation}
\label{sec:quartic:monomial}
\begin{figure}[t]
\begin{center}
\includegraphics[width=0.75\textwidth]{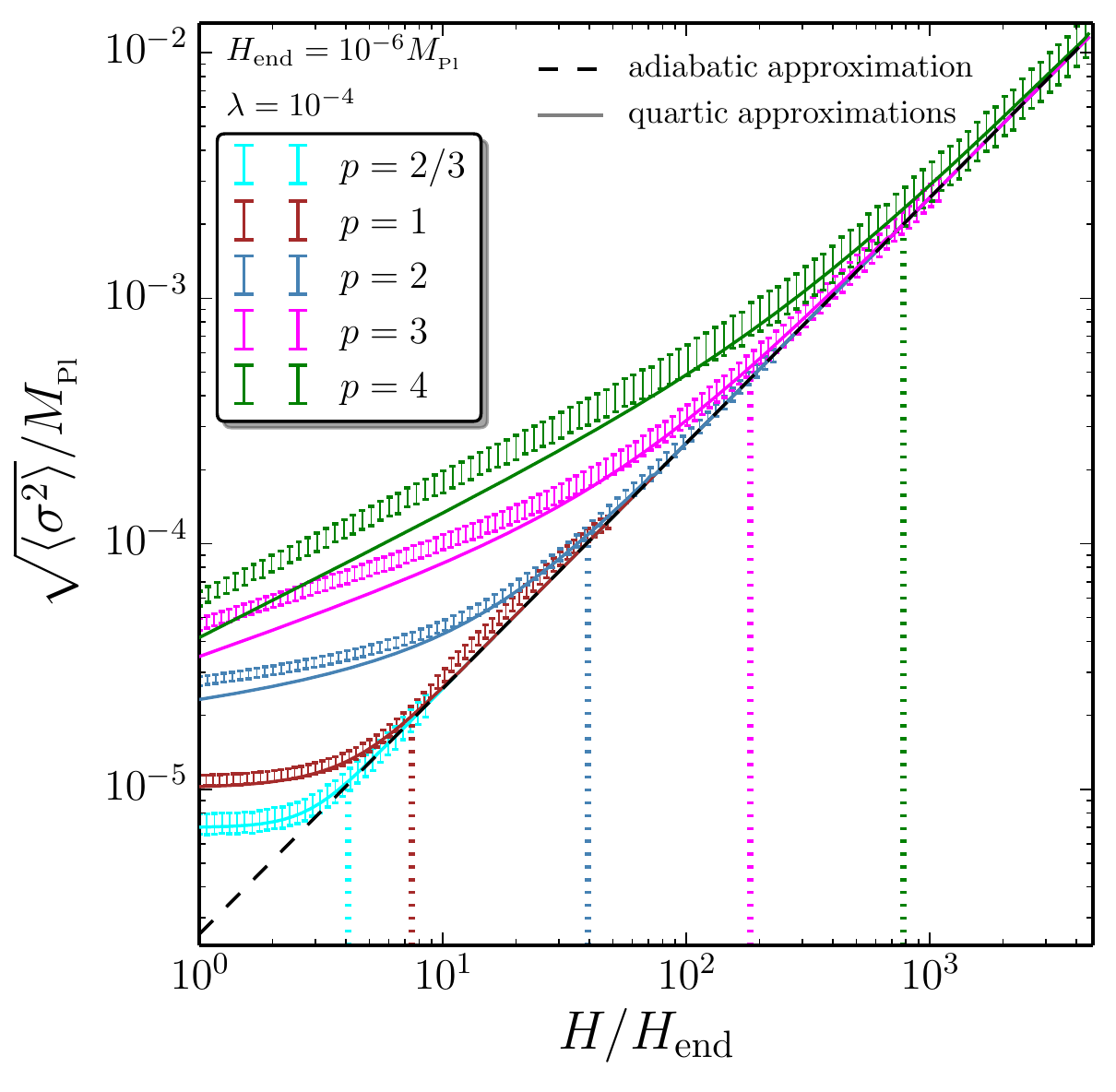}
\caption{Standard deviation $\sqrt{\langle \sigma^2 \rangle}$ of the distribution of a quartic spectator field with potential $V(\sigma)=\lambda\sigma^4$ with $\lambda=10^{-4}$, as a function of time parametrised by the Hubble scale $H$ (time flows from the right to the left). The inflaton potential is of the monomial type $V(\phi)\propto\phi^p$, with $H_\uend=10^{-6}\Mp$. The coloured \textsc{i} symbols correspond to numerical solutions of the Langevin equation where $10^{5}$ realisations of \Eq{eq:Langevin} have been produced for the values of $p$ displayed in the legend. The centres of the vertical bars correspond to ensemble averages of $\sigma^2$ while the heights of the bars are statistical noise estimates (due to having a finite number of realisations only) obtained from the jackknife resampling technique. The realisations are initially drawn according to the adiabatic distribution~(\ref{eq:Pstat}). The black dashed line corresponds to this adiabatic value~(\ref{eq:quartic:deSitter}) for $\langle\sigma^2\rangle$. The coloured dashed vertical lines denote the value of $H$ such that the argument of the Bessel functions in \Eq{eq:analytic_quart} equals one, which corresponds to $H_\mathrm{adiab}$ given by \Eq{eq:quartic:Hadiab} up to an order one prefactor. One can see that when $H$ drops below $H_\mathrm{adiab}$, the numerical solutions depart from the de-Sitter equilibrium, denoting the end of the adiabatic regime. Finally, the coloured solid lines correspond to the quartic approximation~(\ref{eq:analytic_quart}).\label{fig:quartic:sigmaend:time}}
\end{center}
\end{figure}
If the inflaton potential is monomial and of the form $V(\phi)\propto \phi^p$, the Hubble factor is given by \Eq{eq:Hubble} and varies over time scales of order $N_H=(H/H_\uend)^{4/p}$ as explained in \Sec{sec:validity-stat}. Making use of \Eq{eq:quartic:Nrelax}, the adiabatic condition $N_H\gg N_\mathrm{relax}$ then requires $H\gg H_\mathrm{adiab}$, where
\bea
\label{eq:quartic:Hadiab}
H_\mathrm{adiab} \equiv \lambda^{-p/8} H_\uend\, .
\eea
A fundamental difference with the quadratic spectator is that in the quartic case, for all values of $p$, there always exists an adiabatic regime at early times. However, it is not guaranteed  that this regime is consistent with the classical inflaton solution~(\ref{eq:Hubble}), \ie extends beyond the eternal inflationary phase. This is the case only if $H_\mathrm{adiab}<H_\mathrm{eternal}$, where $H_\mathrm{eternal}$ is given in \Eq{eq:Hei:def}, that is to say if $\lambda$ is large enough,
\bea
\label{eq:quartic:adiabatic:condition}
\lambda>\left(\frac{H_\uend}{\Mp}\right)^{\frac{8}{p+2}}\, .
\eea
Let us distinguish the case where this condition is satisfied and one can use the stationary solution~(\ref{eq:Pstat}) to describe the distribution in the adiabatic regime independently of initial conditions, and the case where this is not possible.
\subsubsection{Starting out in the adiabatic regime}
If the condition~(\ref{eq:quartic:adiabatic:condition}) is satisfied, one can set initial conditions for the spectator field  $\sigma$ in the adiabatic regime after the eternal inflationary phase. In \Fig{fig:quartic:sigmaend:time}, we present the results of a numerical integration of the Langevin equation~(\ref{eq:Langevin}) in this case [with the values used for $H_\uend$ and $\lambda$, one can check that \Eq{eq:quartic:adiabatic:condition} is satisfied up to $p=10$]. The values of $H_\mathrm{adiab}$ given by \Eq{eq:quartic:Hadiab} are denoted by the vertical coloured dashed lines. When $H\gg H_\mathrm{adiab}$, the numerical results follow the de-Sitter stationary solution~(\ref{eq:quartic:deSitter}) represented by the black dashed line. When $H$ drops below $H_\mathrm{adiab}$, this is not the case anymore, and the distributions are wider at the end of inflation than the adiabatic approximation would naively suggest.

In this regime, the behaviour of $\langle \sigma^2\rangle$ can in fact still be tracked analytically by making use of the quartic ansatz~(\ref{eq:quartic:ansatz}) introduced in \Sec{sec:quartic:plateau}. Indeed, in the case where $H$ is given by \Eq{eq:Hubble}, one can cast \Eq{eq:quartic:quarticappr:alpha:eom} into a Ricatti equation and in \App{sec:nonlin_drift} it is shown that its solution reads
\bea
 \label{eq:analytic_quart}
\langle \sigma^2 (H)\rangle = \frac{\Gamma\left(\frac{3}{4}\right)}{\Gamma\left(\frac{1}{4}\right)}\sqrt{\frac{3}{2\lambda}} \frac{H^2}{2\pi} \dfrac{K_{\frac{p}{4}+\frac{1}{2}}\left[\frac{p}{4\pi}\sqrt{\frac{\lambda}{6}}\frac{\Gamma \left( \frac{1}{4}\right)}{\Gamma \left( \frac{3}{4}\right)} \left( \frac{H}{H_\uend}\right)^{4/p}\right]}{K_{\frac{p}{4}-\frac{1}{2}}\left[\frac{p}{4\pi}\sqrt{\frac{\lambda}{6}}\frac{\Gamma \left( \frac{1}{4}\right)}{\Gamma \left( \frac{3}{4}\right)} \left( \frac{H}{H_\uend}\right)^{4/p}\right] } \,.
\eea
In this expression, $K$ is a modified Bessel function of the second kind. One can note that the argument of the Bessel functions is directly proportional to $N_H/N_\mathrm{relax}$, confirming that this ratio controls the departure from the adiabatic solution~(\ref{eq:quartic:deSitter}). At early times when $N_H \gg N_\mathrm{relax}$, or equivalently $H\gg H_\mathrm{adiab}$, one can expand the Bessel functions in the large argument limit, $K_\alpha(x)\simeq \sqrt{\pi/(2x)}\ee^{-x}$, and one recovers the adiabatic approximation~(\ref{eq:quartic:deSitter}). The formula~(\ref{eq:analytic_quart}) is displayed in \Fig{fig:quartic:sigmaend:time} with the solid coloured lines. One can see that even when $H<H_\mathrm{adiab}$, it still provides a reasonable approximation to the numerical solutions. One can also notice that the lower $p$ is, the better this quartic approximation. At the end of inflation, $N_H/N_\mathrm{relax} = \sqrt{\lambda} \ll 1$, so the Bessel functions can be expanded in the small argument limit, which depends on the sign of the index of the Bessel function.\footnote
{In the limit $x\ll 1$, if $\alpha<0$, $K_\alpha(x)\simeq \Gamma(-\alpha) 2^{-1-\alpha}x^\alpha$, if $\alpha>0$, $K_\alpha(x)\simeq \Gamma(\alpha) 2^{\alpha-1} x^{-\alpha}$ and if $\alpha=0$, $K_\alpha(x)\simeq \ln(2/x)-\gamma$, where $\gamma\simeq 0.577$ is the Euler constant~\cite{Abramovitz:1970aa}.}
Because the index of the Bessel function in the denominator of \Eq{eq:analytic_quart} is proportional to $p-2$, this leads to different results whether $p$ is smaller or larger than $2$, namely
\bea
\left\langle \sigma_\uend^2\right\rangle \simeq
\left\lbrace
\begin{array}{lcc}
 \dfrac{\Gamma\left(\frac{1}{2}+\frac{p}{4}\right)}{\Gamma\left(\frac{1}{2}-\frac{p}{4}\right)}
\left[\sqrt{\dfrac{3}{2}}\dfrac{\Gamma\left(\frac{3}{4}\right)}{\Gamma\left(\frac{1}{4}\right)}\right]^{1+\frac{p}{2}}
\dfrac{\left(\frac{16\pi}{p}\right)^{\frac{p}{2}}}{2\pi}
\dfrac{H_\uend^2}{\lambda^{\frac{1}{2}+\frac{p}{4}}} & \quad\quad\quad & \mathrm{if}\ p<2\\
& & \\
 \dfrac{6 }{\bar{\gamma}-\ln(\lambda)} \dfrac{\Gamma^2\left(\frac{3}{4}\right)}{\Gamma^2\left(\frac{1}{4}\right)} \dfrac{H_\uend^2}{\lambda}& \quad\quad\quad & \mathrm{if}\ p=2\\
& & \\
 \left(3-\frac{6}{p}\right)\dfrac{\Gamma^2\left(\frac{3}{4}\right)}{\Gamma^2\left(\frac{1}{4}\right)}\dfrac{H_\uend^2}{\lambda}& \quad\quad\quad &\mathrm{if}\ p>2
\end{array}
\right.\quad ,
\label{eq:quartic:sigmaend:adiab}
\eea
where we have defined $\bar{\gamma}\equiv 2\ln[4\pi\sqrt{6}\Gamma(3/4)/\Gamma(1/4)]-2\gamma\simeq 3.53$, where $\gamma$ is the Euler constant. Ignoring the overall constants of order one, if $p\geq 2$, one finds $\langle\sigma_\uend^2\rangle \sim H_\uend^2/\lambda$, and if $p<2$, $\langle\sigma_\uend^2\rangle \sim H_\uend^2/\lambda^{1/2+p/4}$. This needs to be compared to the de-Sitter case~(\ref{eq:quartic:deSitter}) where $\langle\sigma_\uend^2\rangle \sim H_\uend^2/\sqrt{\lambda}$. In monomial inflation,  $\langle\sigma_\uend^2\rangle$ is therefore larger than in plateau inflation for the same value of $H_\uend$, by a factor $\lambda^{-p/4}$ if $p<2$ and $\lambda^{-1/2}$ if $p\geq 2$. One should also note that the condition~(\ref{eq:quartic:adiabatic:condition}) for the adiabatic regime to extend beyond the eternal inflationary phase can be substituted into \Eq{eq:quartic:sigmaend:adiab} and gives rise to $\sqrt{\langle\sigma_\uend^2\rangle}/\Mp \ll (H_\uend/\Mp)^{(p-2)/(p+2)}$ if $p\geq 2$ and $\sqrt{\langle\sigma_\uend^2\rangle}/\Mp \ll 1$ if $p < 2$. In both cases, the spectator field displacement at the end of inflation is therefore sub-Planckian.
\subsubsection{Starting out away from the adiabatic regime}
\begin{figure}[t]
\begin{center}
\includegraphics[width=0.49\textwidth]{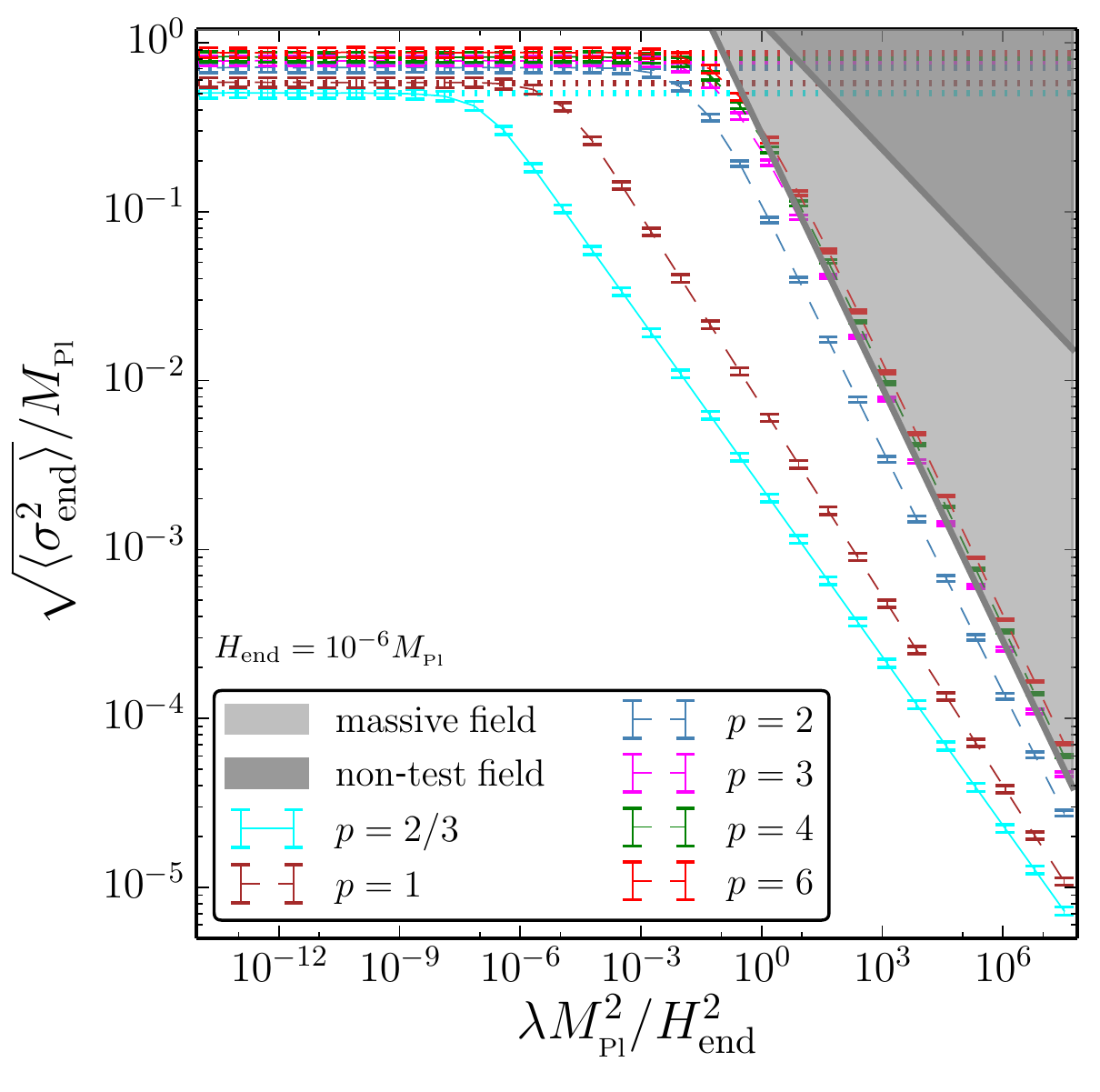}
\includegraphics[width=0.49\textwidth]{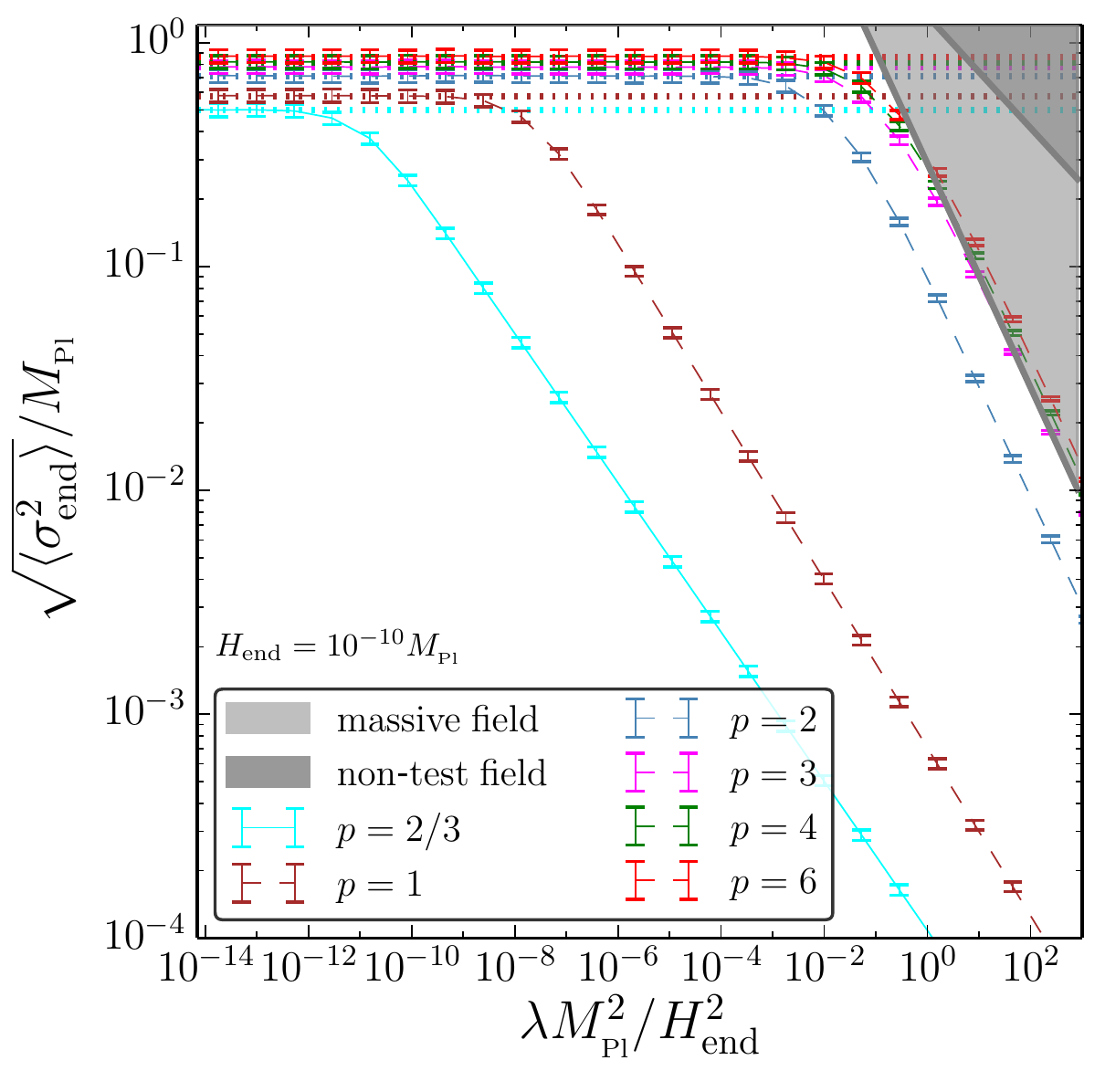}
\caption{Field displacement $\sqrt{\langle \sigma^2 \rangle}$ acquired by a quartic spectator field with potential $V(\sigma)=\lambda\sigma^4$ at the end of inflation, as a function of $\lambda \Mp^2/H_\uend^2$, for $H_\uend=10^{-6}\Mp$ (left panel) and  $H_\uend=10^{-10}\Mp$ (right panel).  The inflaton potential is of the monomial type $V(\phi)\propto\phi^p$. The coloured \textsc{i} symbols correspond to numerical solutions of the Langevin equation where $10^{5}$ realisations of \Eq{eq:Langevin} have been produced for the values of $p$ displayed in the legend. The centres of the vertical bars correspond to ensemble averages of $\sigma^2$ while the heights of the bars are statistical noise estimates (due to having a finite number of realisations only) obtained from the jackknife resampling technique. All realisations are initiated with $\sigma=0$ at $H=H_\mathrm{eternal}$. The horizontal dashed lines correspond to \Eq{eq:quadratic:sigmaend:pgt2} to which the numerical results asymptote in the limit $\lambda\rightarrow 0$. The pale grey region corresponds to $m_\ueff>H$ where the spectator field is not light and our calculation does not apply, and the dark region stands for $\lambda \sigma_\uend^4>3\Mp^2H_\uend^2$ where $\sigma$ cannot be considered as a spectator field anymore.
\label{fig:quartic:sigmaend:nonadiabatic}}
\end{center}
\end{figure}
If the condition~(\ref{eq:quartic:adiabatic:condition}) is not satisfied, the adiabatic regime lies entirely within the eternal inflationary phase and cannot be used to erase initial conditions. In this case, the spectator field displacement at the end of inflation is thus strongly dependent on initial conditions at the start of the classical inflaton evolution. In this section, we derive a lower bound on $\langle\sigma_\uend^2\rangle$, assuming that it vanishes when $H=H_\mathrm{eternal}$ and solving the subsequent stochastic dynamics numerically. The result is presented in \Fig{fig:quartic:sigmaend:nonadiabatic} where $\langle\sigma_\uend^2\rangle$ is displayed as a function of $\lambda\Mp^2/H_\uend^2$ for $H_\uend=10^{-6}\Mp$ (left panel) and for $H_\uend=10^{-10}\Mp$ (left panel). The two cases $p\geq 2$ and $p<2$ must be treated separately.
\paragraph{Case where $p\geq 2$}\mbox{}\\
If $p\geq 2$, it was shown in \Sec{sec:quadratic:pgt2} that a light quadratic spectator field always acquires a super-Planckian field value at the end of inflation. The mean effective mass of the quartic spectator field is given by
 \bea
 \label{eq:quartic:effectivemass}
m_\ueff^2 = 12 \lambda \left\langle \sigma^2 \right\rangle \, ,
 \eea
and is smaller than $H_\uend$ for $\sqrt{\langle\sigma_\uend^2\rangle}\sim \Mp$ if $\lambda<H_\uend^2/\Mp^2$. This explains why, in \Fig{fig:quartic:sigmaend:nonadiabatic}, in the regime $\lambda<H_\uend^2/\Mp^2$, one recovers \Eq{eq:quadratic:sigmaend:pgt2} that is displayed with the horizontal coloured lines, and which shows that the spectator field acquires a super-Planckian field value in this case. Otherwise, if $H_\uend^2/\Mp^2<\lambda<(H_\uend/\Mp)^{8/(p+2)}$ [the upper bound coming from breaking the inequality~(\ref{eq:quartic:adiabatic:condition})], one can see in \Fig{fig:quartic:sigmaend:nonadiabatic} that the field displacement can be made sub-Planckian, but that its effective mass becomes of order $H$.\footnote{Strictly speaking, the present calculation does not apply when the effective mass of the spectator field is of order $H$ or larger. However, if the effects of the mass were taken into account, the amplitude of the noise term in \Eq{eq:Langevin} would not be $H/(2\pi)$ but would become smaller as $m_\ueff$ approaches $H$. This would result in a smaller value for $\langle\sigma^2\rangle$, hence for $m_\ueff$, and therefore a larger noise amplitude. One can expect the two effects to compensate for a value of $m_\ueff$ around $H$.} In this regime, the spectator field cannot be considered as light anymore.
\paragraph{Case where $p< 2$}\mbox{}\\
If $p< 2$, it was shown in \Sec{sec:quadratic:plt2} that a quadratic spectator field acquires a super-Planckian field value at the end of inflation if its mass is smaller than $H_\uend (H_\uend/\Mp)^{(2-p)/(2+p)}$, see \Eq{eq:quadratic:adiabstart:condition}. When evaluated at the Planck scale, the effective mass~(\ref{eq:quartic:effectivemass}) of the quartic spectator field is smaller than this threshold when $\lambda < (H_\uend/\Mp)^{8/(2+p)}$, which exactly corresponds to breaking the inequality~(\ref{eq:quartic:adiabatic:condition}). One can check in \Fig{fig:quartic:sigmaend:nonadiabatic} that when $\lambda < (H_\uend/\Mp)^{8/(2+p)}$, one does indeed recover \Eq{eq:quadratic:sigmaend:pgt2} which is displayed with the horizontal dashed coloured lines. One concludes that in this case, the spectator field always acquires a field value at least of order the Planck mass at the end of inflation.
\mbox{}\\
\mbox{}\\
\indent The situation is summarised in the second line of table~\ref{table:summary} in \Sec{sec:conclusions}. If $\lambda>(H_\uend/\Mp)^{8/(p+2)}$, the spectator field is sub-Planckian at the end of inflation. Otherwise, if $p\geq 2$, either the spectator field is super-Planckian or not light at the end of inflation, and if $p<2$, it is always super-Planckian. Considering the quadratic spectator discussed in \Sec{sec:quad_spec} where it was shown that super-Planckian field displacements are usually generated at the end of inflation, one thus concludes that an additional self-interacting term $\lambda\sigma^4$ in the potential can render the field value sub-Planckian if $\lambda$ is large enough, namely if $\lambda>(H_\uend/\Mp)^{8/(p+2)}$. One can check that for such a value of $\lambda$, if $V(\sigma)=m^2\sigma^2/2+\lambda\sigma^4$ with $m<H_\uend$, the quartic term always dominates over the quadratic one when $\sigma\sim\Mp$, which is consistent.
\section{Axionic spectator}
\label{sec:axion_spec}
In \Sec{sec:quad_spec}, it was shown that quadratic spectator fields with potential $V(\sigma)=m^2\sigma^2/2$ typically acquire super-Planckian field displacements at the end of inflation if the inflaton potential is of the form $V(\phi)\propto\phi^p$ with $p\geq 2$ at large-field value or with $p<2$ and $m/H_\uend<(H_\uend/\Mp)^{(2-p)/(2+p)}$. In \Sec{sec:quart_spec}, we discussed how adding a quartic self-interaction term in the potential could help to tame these super-Planckian values. In this section, we investigate another possibility, which consists in making the field space compact and of sub-Planckian extent. This is typically the case for axionic fields, with periodic potentials of the type
\bea
\label{eq:axionpot}
V(\sigma) = \Lambda^4\left[ 1- \cos \left( \frac{\sigma}{f} \right)\right]\,.
\eea
In this expression, $\Lambda$ and $f$ are two mass scales that must satisfy $\Lambda^2<fH_\uend$ in order for the curvature of the potential to remain smaller than the Hubble scale throughout inflation, \ie for the axionic field to remain light, which we will assume in the following.
\subsection{Plateau inflation}
\label{sec:axionic:plateau}
As explained in \Sec{sec:validity-stat}, if the inflaton potential is of the plateau type, $H$ can be approximated by a constant and the spectator field value reaches the de-Sitter equilibrium~(\ref{eq:Pstat}). If $H\gg \Lambda$, such a distribution is approximately flat, in which case $\langle\sigma^2\rangle\simeq \pi^2 f^2/3$ if $\sigma$ is restricted to one period of the potential~(\ref{eq:axionpot}). In this regime, the classical drift due to the potential gradient in \Eq{eq:Langevin} can be neglected and the spectator field experiences a free diffusion process. The relaxation time is therefore the time it takes to randomise $\sigma$ over the period of the potential and is given by $N_\mathrm{relax}\simeq (\pi^2f/H)^2$. In the opposite limit when $H\ll \Lambda$, the distribution is localised close to the minimum of the potential where it can be approximated by a quadratic function $V(\sigma)\simeq m^2\sigma^2/2$ with mass $m^2=\Lambda^4/f^2$. In this case, according to \Sec{sec:plateau_quad_spec}, one has $\langle \sigma^2 \rangle = 3H^4f^2/(8\pi^2\Lambda^4)$, and the relaxation time is of order $N_\mathrm{relax}=H^2/m^2\simeq H^2f^2/\Lambda^4$.
\subsection{Monomial inflation}
\label{sec:axionic:monomial}
\begin{figure}[t]
\begin{center}
\includegraphics[width=0.75\textwidth]{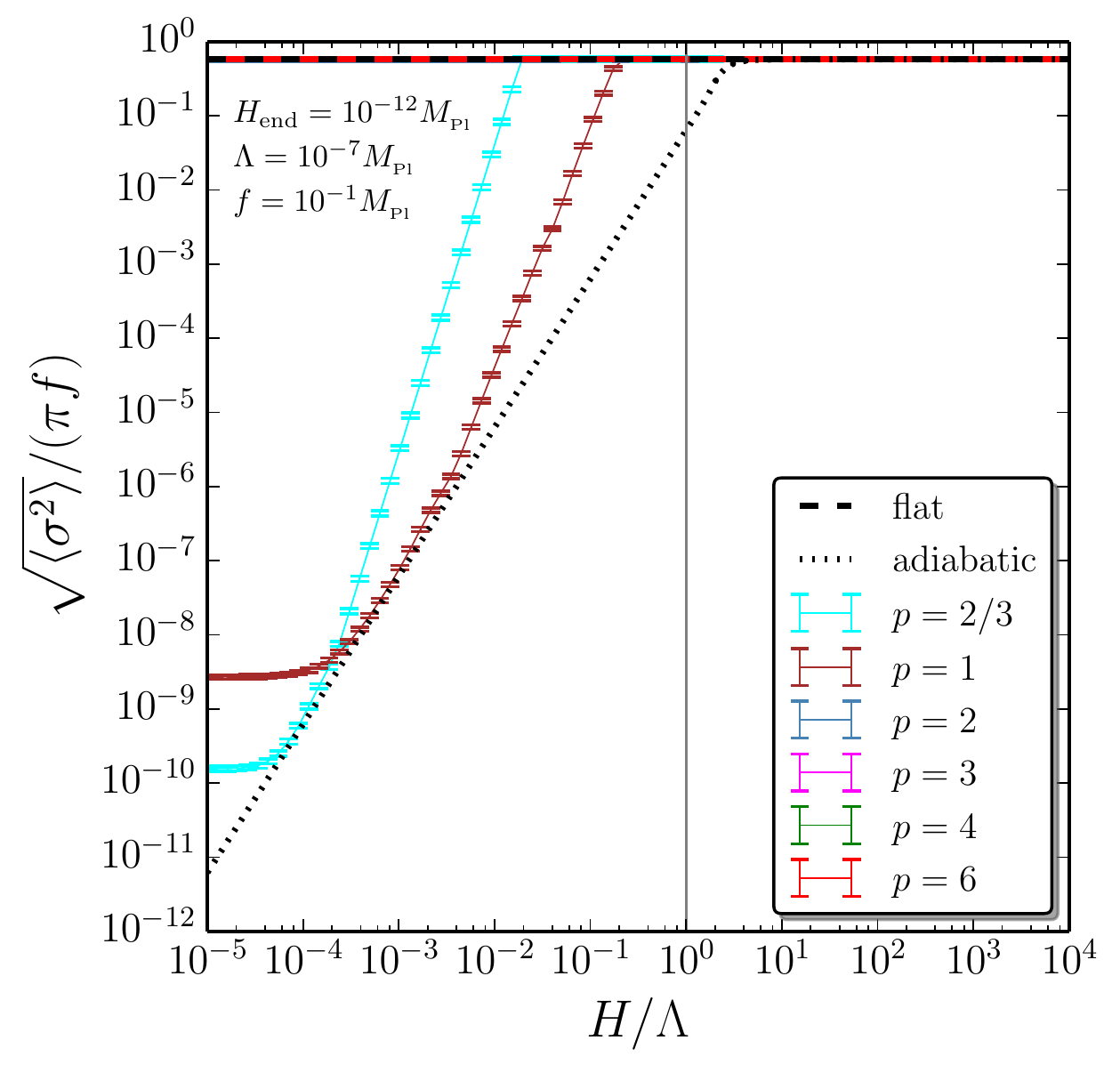}
\caption{Standard deviation $\sqrt{\langle \sigma^2 \rangle}$ of the distribution of an axionic spectator field with potential $V(\sigma)=\Lambda^4[1-\cos(\sigma/f)]$ with $\Lambda=10^{-7}\Mp$ and $f=10^{-1}\Mp$, as a function of time parametrised by the Hubble scale $H$ (time flows from the right to the left). The inflaton potential is of the monomial type $V(\phi)\propto\phi^p$, with $H_\uend=10^{-12}\Mp$. The coloured \textsc{i} symbols correspond to numerical solutions of the Langevin equation where $10^{6}$ realisations of \Eq{eq:Langevin} have been produced for the values of $p$ displayed in the legend. The centres of the vertical bars correspond to ensemble averages of $\sigma^2$ while the heights of the bars are statistical noise estimates (due to having a finite number of realisations only) obtained from the jackknife resampling technique. The realisations are initially drawn according to a flat distribution when $H/\Lambda=10^4$. The black dashed line  corresponds to the standard deviation of a distribution that is flat over one period of the potential, $\langle \sigma^2\rangle=\pi^2f^2/3$.  The black dotted line corresponds to the adiabatic solution~(\ref{eq:Pstat}), which remains flat when $H$ is larger than $\Lambda$, represented by the grey vertical line. When $p\geq 2$ the distributions remain flat until the end of inflation (and one cannot distinguish the different values of $p$ that are superimposed). When $p<2$, the distributions narrow down once $H\ll \Lambda$ since the parameters have been chosen to satisfy \Eq{eq:axionic:adiabatic:condition}.
\label{fig:axionic:sigmaend:time}}
\end{center}
\end{figure}

If inflation is realised by a monomial potential $V(\phi)\propto \phi^p$, there is always an epoch when $H>\Lambda$ in the past and during which the spectator field distribution is made flat within a number of \efolds of order $N_\mathrm{relax}\simeq (\pi^2f/H)^2$. Therefore, contrary to the quadratic and to the quartic spectators, the field displacement of an axionic spectator at the end of inflation is always independent of initial conditions, provided that inflation lasts long enough. If $\Lambda<H_\uend$, the distribution remains flat until the end of inflation. In the opposite case, when $H$ drops below $\Lambda$, the subsequent dynamics of $\sigma$ depends on whether $p\geq 2$ or $p<2$.
\subsubsection{Case where $p\geq 2$}
\label{sec:axionic:pgt2}
If $p\geq 2$, in \Sec{sec:quad_spec} it was shown that the evolution of a quadratic field with mass $m<H_\uend$ is effectively described by a free-diffusion process where the potential drift can be neglected. For an axionic spectator, the potential is always flatter than its quadratic expansion around its minimum and can therefore also be neglected. As a consequence, the distribution remains flat until the end of inflation and one finds $\langle\sigma_\uend^2\rangle\simeq \pi^2 f^2/3$.
\subsubsection{Case where $p< 2$}
\label{sec:axionic:plt2}
If $p<2$, in \Sec{sec:quad_spec} it was shown that the distribution of a quadratic field with mass $m<H_\uend$ tracks the adiabatic equilibrium until $H=H_\mathrm{adiab}$, where $H_\mathrm{adiab}$ is given by \Eq{eq:Hadiab:def}, and remains frozen afterwards. This implies that an axionic spectator distribution narrows down from a flat profile if $H_\mathrm{adiab}<\Lambda$, which gives rise to
\bea
\label{eq:axionic:adiabatic:condition}
\frac{\Lambda}{H_\uend}>\left(\frac{f}{H_\uend}\right)^{\frac{p}{p+2}}\, .
\eea
Notice that for this condition to be compatible with the light-field prescription given below \Eq{eq:axionpot}, one must have $H_\uend<f$ for $p<2$ (which makes sense, otherwise the distribution would be randomised over one \efold~even towards the end of inflation). In this case, $\langle \sigma^2\rangle$ settles down to $3 H_\mathrm{adiab}^4/(8\pi^2 m^2)$, which gives rise to
\bea
\label{eq:axionic:adiabatic:sigmaend}
\sqrt{\left\langle \sigma_\uend^2 \right\rangle} \simeq \sqrt{\frac{3}{2}}\frac{H_\uend}{2\pi}\left(\frac{H_\uend f}{\Lambda^2}\right)^{\frac{2+p}{2-p}}\, .
\eea
If \Eq{eq:axionic:adiabatic:condition} is not satisfied however, the field distribution remains flat until the end of inflation and one has $\langle\sigma_\uend^2\rangle\simeq \pi^2 f^2/3$.
\mbox{}\\
\mbox{}\\
\indent
In order to check the validity of these considerations, in \Fig{fig:axionic:sigmaend:time} we present numerical solutions of the Langevin equation~(\ref{eq:Langevin}). When $p\geq2$, one can check that the distributions remain flat until the end of inflation. The values of the parameters $\Lambda$, $f$ and $H_\uend$ have been chosen to satisfy \Eq{eq:axionic:adiabatic:condition}, which explains why for $p<2$, the distributions narrow down once $H$ drops below $\Lambda$ (otherwise, we have checked that even when $p<2$, the distributions remain flat). However, one can see that when the distributions start moving away from the flat configuration, they do not exactly follow the adiabatic solution displayed with the black dotted line, even though $H>H_\mathrm{adiab}$. This is because in the above discussion, we have approximated the axionic potential with its quadratic expansion around its minimum, which is not strictly valid at the stage where the distribution is still flat and sensitive to the full potential shape. Nonetheless, the distributions converge towards the adiabatic profile at later time and the final value of $\langle \sigma^2 \rangle$ is well described by \Eq{eq:axionic:adiabatic:sigmaend}.

The situation is summarised in the third line of table~\ref{table:summary} in \Sec{sec:conclusions}. If $H_\uend > \Lambda$, $p\geq 2$, or $p<2$ with $\Lambda<H_\uend (f/H_\uend)^{p/(p+2)}$, the distribution of the axionic spectator remains flat until the end of inflation and $\sqrt{\smash[b]{\langle\sigma_\uend^2}\rangle }= \pi f/\sqrt{3}$. Only if $p<2$ with $\Lambda>H_\uend (f/H_\uend)^{p/(p+2)}$ does the distribution narrow down and $\sqrt{\smash[b]{\langle\sigma_\uend^2\rangle}} \simeq H_\uend(H_\uend f /\Lambda^2)^{(2+p)/(2-p)} $. In all cases, if $f$ is sub-Planckian, the typical field displacement obviously remains sub-Planckian as well.
\section{Information retention from initial conditions}
\label{sec:sigmaendpriors}
When calculating the field value acquired by spectator fields at the end of inflation, we have found situations in which initial conditions are erased by the existence of an adiabatic regime at early times, and situations in which this is not the case. In this section, we propose to quantify this memory effect using information theory in order to better describe the amount of information about early time physics (potentially pre-inflationary) available in the final field displacements of spectator fields. The relative information between two distributions $P_1(\sigma)$ and $P_2(\sigma)$ can be measured using the Kullback-Leibler divergence~\cite{kullback1951} $\DKL$,
\bea
\DKL\left(P_1 \vert\vert P_2 \right) \equiv \int^{\infty}_{-\infty} {P_1}\left(\sigma \right) \log_2 \left[\frac{{P_1}\left(\sigma \right)}{P_2 \left(\sigma \right)} \right] \dd \sigma\, .
\label{eq:DKL}
\eea
It is invariant under any reparametrisation $\sigma^\prime = f(\sigma)$, and since it uses a logarithmic score function as in the Shannon's entropy, it is a well-behaved measure of information~\cite{bernardo:2008}. Considering two initial distributions separated by an amount of information $\delta\DKL^0$, giving rise to two final distributions separated by $\delta\DKL^\uend$, we define the information retention criterion by
\bea
\label{eq:calI:def}
\mathcal{I}\equiv\frac{\delta\DKL^\uend}{\delta\DKL^0}\, .
\eea
When $\mathcal{I}<1$, the initial information is contracted by the dynamics of the distributions. This is typically the case when there is an attractor, or an adiabatic regime, which tends to erase the initial conditions dependence of final states. When $\mathcal{I}>1$, the initial information is amplified and the final state is sensitive to initial conditions. Values of $\mathcal{I}\gg 1$ might signal the presence of chaotic dynamics in which case initial conditions are difficult to infer. For this reason, $\mathcal{I}=\order{1}$ represents an optimal situation in terms of initial conditions reconstruction. In practice, $\mathcal{I}$ depends both on the initial (or final) state around which the infinitesimal variation is performed, and on the direction in the space of distributions along which it is performed.
\begin{figure}
\centering
\includegraphics[width=0.49\textwidth]{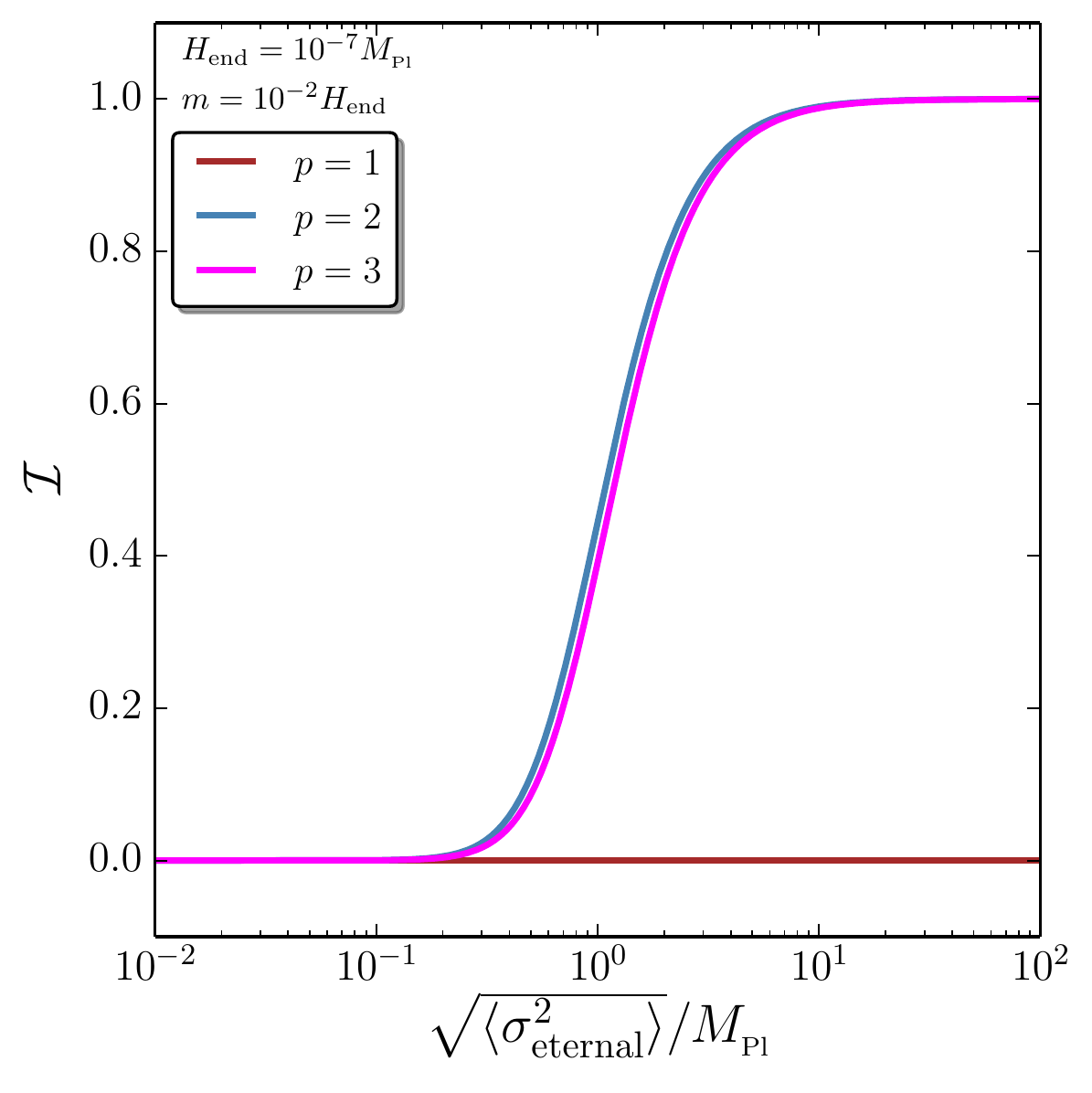}
\includegraphics[width=0.49\textwidth]{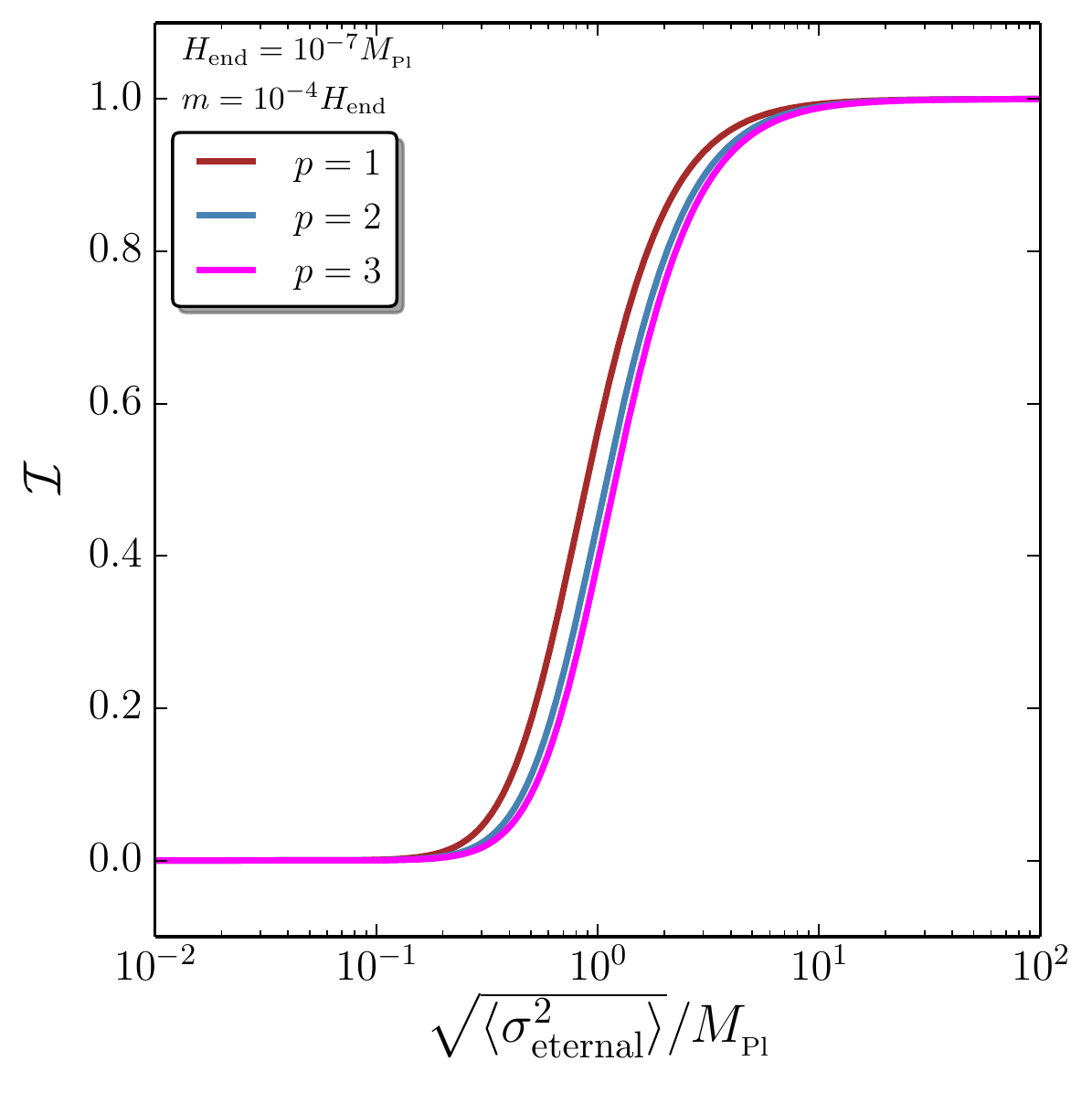}
\caption{\label{fig:quadratic_information} Information retention criterion~(\ref{eq:calI:def}) as a function of the initial standard deviation $\sqrt{\langle\sigma_\mathrm{eternal}^2\rangle}$ for a quadratic spectator field with potential $V(\sigma)=m^2\sigma^2/2$, if inflation is driven by a monomial potential $V(\phi)\propto\phi^p$. Initial conditions are set at $H_0=H_\mathrm{eternal}$ where the inflaton exits the eternal inflationary regime. In both panels, $H_\uend=10^{-7}\Mp$, and $m=10^{-2} H_\uend$ in the left panel and $m=10^{-4}H_\uend$ in the right panel. Different colours represent different values of $p$. If $p\geq 2$, initial conditions are not erased but provide a subdominant contribution to the final distribution if the field displacement is initially sub-Planckian. This is why, if $\sqrt{\smash[b]{\langle \sigma^2_\mathrm{eternal}\rangle}} \ll \Mp$, $\mathcal{I}\simeq 0$, while if $\sqrt{\smash[b]{\langle \sigma^2_\mathrm{eternal}\rangle}} \gg \Mp$, $\mathcal{I}\simeq 1$. In the left panel, the condition~(\ref{eq:quadratic:adiabstart:condition}) is satisfied for $p=1$, so initial conditions are erased ($\mathcal{I}\simeq 0$), while in the right panel, the condition~(\ref{eq:quadratic:adiabstart:condition}) is not satisfied for $p=1$ which therefore behaves as the cases $p\geq 2$.}
\end{figure}

For concreteness, let us restrict the analysis to the space of symmetric Gaussian distributions, fully characterised by a single parameter, $\langle \sigma^2\rangle$. In this case, \Eq{eq:DKL} gives rise to
\bea
\label{eq:DKL:Gaussian}
\DKL\left(P_1 \vert\vert P_2 \right) = \frac{1}{2\ln 2}\left[\frac{\left\langle \sigma_2^2\right\rangle}{\left\langle \sigma_1^2\right\rangle}-\ln\left(\frac{\left\langle \sigma_2^2\right\rangle}{\left\langle \sigma_1^2\right\rangle}\right)-1\right]\, ,
\eea
where $\langle\sigma_1^2\rangle$ (respectively $\langle\sigma_2^2\rangle$) is the variance of $P_1$ (respectively $P_2$). One then has $\delta\DKL  = (\delta\langle\sigma^2\rangle/\langle\sigma^2\rangle)^2/(4\ln 2)$, which gives rise to\footnote
{The same expression is obtained if one uses the Jensen-Shannon divergence as a measure of the relative information between two distributions,
\bea
\label{eq:jsd}
\DJS\left(P_1 \vert\vert P_2 \right) =  \frac{1}{2}\DKL \left(P_1\left\vert\left\vert \frac{P_1+P_2}{2}\right.\right.\right) + \frac{1}{2}\DKL \left(P_2\left\vert\left\vert \frac{P_1+P_2}{2} \right.\right.\right)  \,,
\eea
which is a symmetrised and smoothed version of the Kullback-Leibler divergence. The Jensen-Shannon divergence between two Gaussian distributions cannot be expressed in a closed form comparable to \Eq{eq:DKL:Gaussian}. However, in the limit where the two Gaussian distributions have variances $\langle\sigma^2\rangle$ and $\langle\sigma^2\rangle+\delta \langle\sigma^2\rangle$ infinitesimally close one to the other, one can expand the integrands of \Eq{eq:jsd} at quadratic order in $\delta \langle\sigma^2\rangle$ and obtain $\delta \DJS = (\delta\langle\sigma^2\rangle/\langle\sigma^2\rangle)^2/(16\ln 2)= \delta\DKL/4$. As a consequence, $\delta \DJS^\uend/\delta\DJS^0 = \delta \DKL^\uend/\delta\DKL^0$ and the same information retention criterion is obtained.
}
\bea
\label{eq:info-ret:Gaussian}
\mathcal{I} = \left(  \frac{\partial\ln \langle \sigma^2_\uend\rangle}{\partial  \ln\langle \sigma^2_{0}\rangle} \right)^2\,.
\eea
In practice, the functional relationship between $\langle\sigma_0^2\rangle$ and $\langle\sigma_\uend^2\rangle$ depends on the details of the stochastic dynamics followed by $\sigma$. When $\langle \sigma_\uend^2\rangle$ is independent of $\langle \sigma_0^2\rangle$ for instance, initial conditions are irrelevant to determine the final state and $\mathcal{I}=0$.

For quadratic spectator fields, in \Sec{sec:quad_spec} it was shown that the distributions remain Gaussian if they were so initially, and the relationship~(\ref{eq:meansigmasquare:final}) between $\langle\sigma_0^2\rangle$ and $\langle\sigma_\uend^2\rangle$ was derived. The formula~(\ref{eq:info-ret:Gaussian}) can therefore directly be evaluated, and it is displayed in \Fig{fig:quadratic_information} in the case where inflation is driven by a monomial potential $V\propto\phi^p$ and initial conditions are taken at the time when the inflaton exits the eternal inflationary epoch. When $p\geq 2$, there is no adiabatic regime and therefore no erasure of initial conditions. Since quantum diffusion contributes a field displacement of order the Planck mass, if the initial field value is much smaller than the Planck mass, it provides a negligible contribution to the final field value and one has $\mathcal{I}\simeq 0$. If it is much larger than the Planck mass it provides the dominant contribution to the final field value and  $\mathcal{I}\simeq 1$. In the left panel, the value of $m$ has been chosen so that the condition~(\ref{eq:quadratic:adiabstart:condition}) is satisfied for $p=1$. In this case, initial conditions are erased during the adiabatic regime and one has $\mathcal{I}\simeq 0$. In the right panel, the value chosen for $m$ is such that \Eq{eq:quadratic:adiabstart:condition} is not satisfied and the situation for $p=1$ is similar to the cases $p\geq2$.

For quartic spectator fields, in \Sec{sec:quart_spec} it was shown that either the condition~(\ref{eq:quartic:adiabatic:condition}) is satisfied and initial conditions are erased during an early adiabatic phase, leading to $\mathcal{I}\simeq 0$; or if the condition~(\ref{eq:quartic:adiabatic:condition}) is not satisfied, the dynamics of the spectator field is described by a free diffusion process and the situation is the same as in the right panel of  \Fig{fig:quadratic_information}.

For axionic spectator fields finally, in  \Sec{sec:axion_spec}, initial conditions were shown to always be erased at early times, yielding $\mathcal{I}\simeq 0$.

The amount of information one can recover about the initial state from the final one therefore depends both on the potential of the spectator field and on the inflationary background. Let us stress that in some situations, initial conditions are not erased ($\mathcal{I}\simeq 1$). This suggests that, if observations yield non-trivial constraints on spectator field values at the end of inflation in our local patch, one may be able to infer a non-trivial probability distribution on its field value at much earlier time, for instance when one leaves the regime of eternal inflation. This might be relevant to the question~\cite{Linde:2005yw} of whether observations can give access to scales beyond the observational horizon.
\section{Conclusion}
\label{sec:conclusions}
The typical field value acquired by spectator fields during inflation is an important parameter of many post-inflationary physical processes. Often, in slow-roll inflationary backgrounds, it is estimated using the stochastic equilibrium solution in de-Sitter space-times~(\ref{eq:Pstat}), since slow  roll is parametrically close to de Sitter. However, slow roll only implies that the Hubble scale $H$ varies over time scales larger than one \efold. Since the relaxation time of a spectator field distribution towards the de-Sitter equilibrium is typically much larger than an \efold, this does not guarantee that the spectator distribution adiabatically tracks the de-Sitter solution. In practice, we have found that when the inflaton potential is monomial at large-field values, the de-Sitter approximation is never a reliable estimate of the spectator typical field value at the end of inflation. Instead, spectator fields acquire field displacements that depend on the details of both the spectator potential and the inflationary background. These results are summarised in table~\ref{table:summary}.

In some cases, the existence of an adiabatic regime at early times leads to an erasure of initial conditions and the spectator field distribution is fully determined by the microphysical parameters of the model. When this is the case, we have showed that spectator fields always acquire sub-Planckian field values at the end of inflation. However, it can also happen that adiabatic regimes either do not exist or take place at a stage where quantum corrections to the inflaton dynamics are large and our calculation does not apply. In such cases, a dependence on  the initial conditions is unavoidable, which we have quantified in the context of information theory. This suggests that observations might have the potential to give access to scales beyond the observable horizon, through processes that are integrated over the whole inflationary period, such as spectator field displacements.

In general, we have found that light spectator fields acquire much larger field displacements during inflation than the de-Sitter approximation suggests, which has important consequences. As an illustration, let us mention one of the curvaton models which is favoured by observations, where inflation is driven by a quartic potential in the presence of a quadratic spectator field, the curvaton, that later dominates the energy budget of the Universe and provides the main source of cosmological perturbations. In order for this model to provide a good fit to the data, the field value of the curvaton at the end of inflation should lie in the range~\cite{Vennin:2015vfa, Vennin:2015egh} $\Gamma_\sigma/\Gamma_\phi \ll \sigma_\uend/\Mp \ll 1$, where $\Gamma_\phi$ and $\Gamma_\sigma$ are the decay rates of the inflaton and of the curvaton, respectively. In this case however, we have found that if inflation starts from the eternal inflation regime, then the curvaton typically acquires a super-Planckian field value at the end of inflation, which challenges this model, at least in its simplest form. As shown in this work, a possible solution could be to add a quartic coupling term to the curvaton potential or to consider axionic curvaton potentials. Whether the model is still in agreement with the data in this case is an important question that we plan to study in a future work.

\newgeometry{top=3.1cm}
\begin{landscape}
\begin{table}
\begin{center}
\includegraphics[height=0.57\textheight]{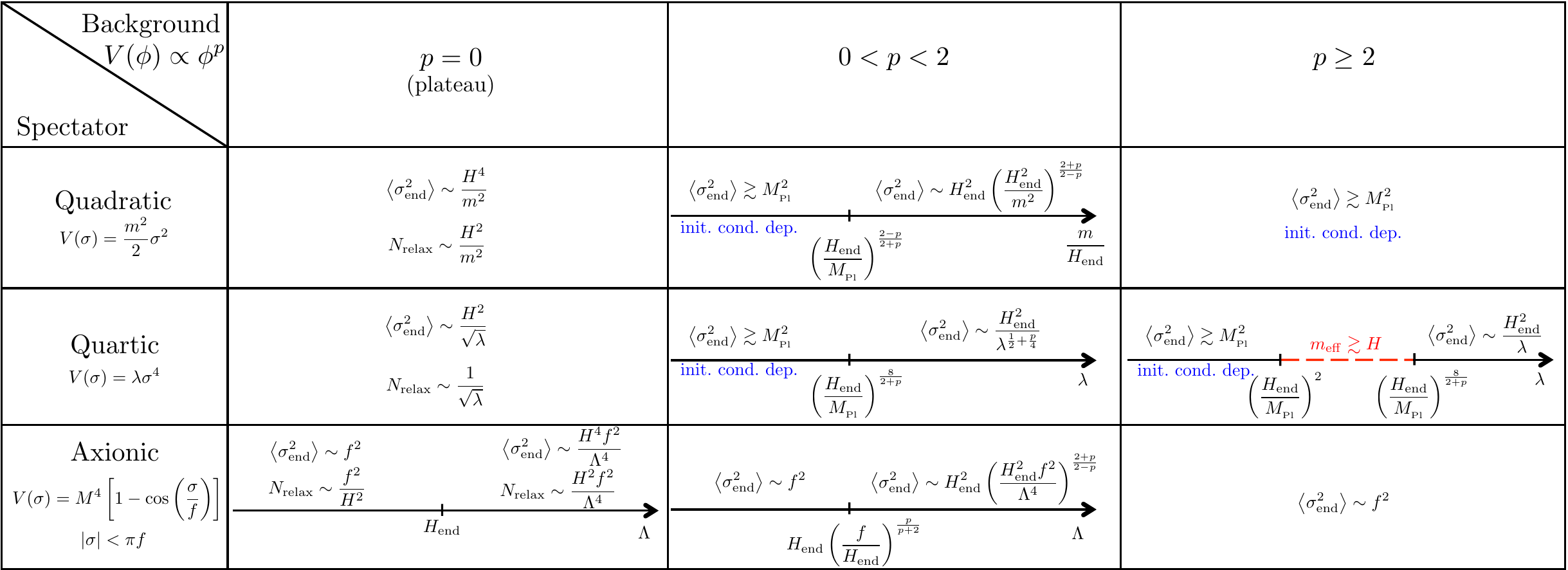}
\caption{Summary of the results obtained in this paper. The stochastic dynamics of spectator scalar fields with quadratic, quartic and axionic potentials have been studied in inflationary backgrounds driven by plateau and monomial potentials. In each case, the typical field displacement $\langle \sigma_\uend^2\rangle$ acquired by spectator fields at the end of inflation is given in this table. When inflation is realised with a plateau potential, the de-Sitter equilibrium is reached within a number of \efolds $N_\mathrm{relax}$ also given in the table. If the inflaton potential receives monomial corrections at large-field values, the de-Sitter approximation is never a reliable estimate of the spectator typical field value, and the result depends on the details of both the spectator potential and the inflationary background. In some cases, the lack of adiabatic attractors at early time also introduces initial conditions dependence (denoted by ``init. cond. dep''.).    }
\label{table:summary}
\end{center}
\end{table}
\end{landscape}
\restoregeometry

\section*{Acknowledgements}
This work was supported by STFC grants ST/N000668/1, ST/K502248/1 and ST/K00090X/1. Numerical computations were done on the Sciama High Performance Compute (HPC) cluster which is supported by the ICG, SEPNet and the University of Portsmouth. C.B. is supported by a Royal Society University Research Fellowship. J.T. acknowledges support from the European Research Council under the European Union's Seventh Framework Programme (FP/2007-2013) / ERC Grant Agreement No. [616170].

\appendix
\section{Statistical moments of quadratic spectators}
\label{sec:quadmoments}
In this section, we derive the first two statistical moments of quadratic spectator fields, for which $V(\sigma)=m^2\sigma^2/2$. If the initial distribution is Gaussian, it remains so throughout the entire evolution so these two moments fully characterise the distribution at any time. Otherwise, higher-order moments can be derived along the same lines.

The first moment can be obtained by taking the stochastic average of \Eq{eq:Langevin}, which gives rise to
\bea
\label{eq:ode:meansigma}
\frac{\dd\langle \sigma \rangle }{\dd N}=-\frac{m^2}{3H^2}\langle \sigma \rangle\, .
\eea
In this expression, the fact that $\sigma$ is a test field plays an important role since it implies that $H$ does not depend on $\sigma$ and is thus a classical (\ie non-stochastic) quantity. Interestingly, \Eq{eq:ode:meansigma} is the same as \Eq{eq:Langevin} in the absence of quantum diffusion, which is why $\langle\sigma\rangle$ follows the classical dynamics
\bea
\label{eq:meansigma:App}
\langle\sigma\left(N\right)\rangle = \langle\sigma\left(N_0\right)\rangle\exp\left[-\frac{m^2}{3}\int_{N_0}^N\frac{\dd{N}^\prime}{H^2({N}^\prime)}\right]\,,
\eea
where $\langle\sigma\left(N_0\right)\rangle$ is the value of $\langle \sigma \rangle$ at the initial time $N_0$.

The second moment can be obtained by multiplying \Eq{eq:Langevin} by $\sigma$ and taking the stochastic average, which leads to
\bea
\label{eq:meansigma2:eom}
\frac{1}{2}\frac{\dd\langle\sigma^2\rangle}{\dd N} = -\frac{m^2}{3H^2}\langle\sigma^2\rangle+\frac{H}{2\pi}\langle \sigma\xi\rangle,
\eea
where $\langle \sigma\xi\rangle$ needs to be calculated separately. This can be done by noticing that a formal solution to \Eq{eq:Langevin} is given by
\bea
\sigma=\int_{A}^N\dd N^\prime \frac{H(N^\prime)}{2\pi}\xi(N^\prime)\exp\left[\int_N^{N^\prime}\frac{m^2}{3H^2(N^{\prime\prime})}\dd N^{\prime\prime}\right]\, ,
\eea
where $A$ is an integration constant. This gives rise to
\begin{align}
\left\langle\sigma(N)\xi(N)\right\rangle& =\int_{A}^N\dd N^\prime \frac{H(N^\prime)}{2\pi}\left\langle\xi(N)\xi(N^\prime)\right\rangle\exp\left[\int_N^{N^\prime}\frac{m^2}{3H^2(N^{\prime\prime})}\dd N^{\prime\prime}\right]\\
& =\int_{A}^N\dd N^\prime \frac{H(N^\prime)}{2\pi}\delta(N-N^\prime)\exp\left[\int_N^{N^\prime}\frac{m^2}{3H^2(N^{\prime\prime})}\dd N^{\prime\prime}\right]\\
&=\frac{1}{2} \frac{H(N)}{2\pi}\, ,
\end{align}
where the factor $1/2$ comes from the fact that the delta function is centred at one of the boundaries of the integral [recall that $\int_{x_0}^{x_1}f(x)\delta(x-x_0)=f(x_0)/2$]. One can then write \Eq{eq:meansigma2:eom} as
\bea
\label{eq:ode:meansigmasquare}
\frac{1}{2}\frac{\dd\langle\sigma^2\rangle}{\dd N} = -\frac{m^2}{3H^2}\langle\sigma^2\rangle+\frac{H^2}{8\pi^2}.
\eea
This equation can be solved and one obtains
\bea
\label{eq:meansigmasquare}
\left\langle \sigma^2(N)\right\rangle = \int_B^N \frac{\dd N^\prime}{4\pi^2} H^2(N^\prime)\exp\left[\frac{2m^2}{3}\int_N^{N^\prime}\frac{\dd N^{\prime\prime}}{H^2(N^{\prime\prime})}\right]\, .
\eea
In this expression, $B$ is an integration constant that can be solved requiring that $\langle \sigma^2\rangle =\langle \sigma^2(N_0)\rangle $ at the initial time $N_0$. This gives rise to
\bea
\label{eq:meansigmasquare:final:App}
\left\langle \sigma^2(N)\right\rangle =& \left\langle \sigma^2(N_0)\right\rangle \exp\left[-\frac{2m^2}{3}\int_{N_0}^{N}\frac{\dd N^{\prime}}{H^2(N^{\prime})}\right]
\\ & 
+ \int_{N_0}^N \dd N^\prime \frac{H^2(N^\prime)}{4\pi^2} \exp\left[\frac{2m^2}{3}\int_N^{N^\prime}\frac{\dd N^{\prime\prime}}{H^2(N^{\prime\prime})}\right]\, .
\eea
In this expression, the structure of the first term is similar to the first moment~(\ref{eq:meansigma:App}), so that the variance of the distribution $\langle \sigma^2 \rangle-\langle\sigma\rangle^2$ evolves according to the same formula as the second moment [\ie one can replace $\langle \sigma^2\rangle$ by $\langle \sigma^2 \rangle-\langle\sigma\rangle^2$ in \Eq{eq:meansigmasquare:final:App} and the formula is still valid].
\section{Adiabatic solution for quartic spectators}
\label{sec:nonlin_drift}
For quartic spectator fields, the Langevin equation is not linear anymore and cannot be solved analytically. In this section we provide a solution using the ansatz
\bea
\label{eq:quartic:ansatz:App}
P(\sigma,N)=
\frac{2 \alpha^{1/4}(N)}{\Gamma\left(\frac{1}{4}\right)}\exp\left[-\alpha(N)\sigma^4\right]\, .
\eea
This ansatz is satisfied by the de-Sitter equilibrium~(\ref{eq:Pstat}), so we expect the solution to be valid at least in the adiabatic regime and potentially beyond. By plugging \Eq{eq:quartic:ansatz:App} into \Eq{eq:FP}, one obtains
\bea
\label{eq:quartic:App:interm2}
  \left( \frac{1}{4\alpha} - \sigma^4 \right) \frac{\dd \alpha}{\dd N} P\left(\sigma,N\right) = \left( \frac{4\lambda}{H^2} - \frac{3H^2\alpha}{2\pi^2} \right) \sigma^2 P\left(\sigma,N\right) + \left( \frac{2H^2\alpha^2}{\pi^2} - \frac{16\lambda\alpha}{3 H^2} \right) \sigma^6 P\left(\sigma,N\right).
\eea
Multiplying this equation by $\sigma^2$ and integrating over $\sigma$, this gives rise to
\bea
\label{eq:quartic:App:interm}
  \left( \frac{\left\langle\sigma^2\right\rangle}{4\alpha} - \left\langle\sigma^6\right\rangle \right) \frac{\dd \alpha}{\dd N}  = \left( \frac{4\lambda}{H^2} - \frac{3H^2}{2\pi^2}\alpha \right) \left\langle\sigma^4\right\rangle + \left( \frac{2H^2}{\pi^2} \alpha^2 - \frac{16\lambda}{3 H^2}\alpha \right) \left\langle\sigma^8\right\rangle P\left(\sigma,N\right) \,.
\eea
From the ansatz~(\ref{eq:quartic:ansatz:App}), the moments $\langle \sigma^2 \rangle$, $\langle \sigma^4 \rangle$, $\langle \sigma^6 \rangle$ and $\langle \sigma^8 \rangle$ are directly related to $\alpha$, through
\bea
\label{eq:moments}
\langle \sigma^2 \rangle = \frac{\Gamma \left( \frac{3}{4}\right) }{\alpha^{1/2}\Gamma \left( \frac{1}{4}\right)}\,,  \qquad
\langle \sigma^4 \rangle = \frac{1}{4\alpha}\,, \qquad
\langle \sigma^6 \rangle = \frac{3\Gamma \left( \frac{3}{4}\right) }{4\alpha^{3/2}\Gamma \left( \frac{1}{4}\right)} \,,  \qquad
\langle \sigma^8 \rangle = \frac{5}{16\alpha^2}\,.
\eea
By substituting these expressions into \Eq{eq:quartic:App:interm}, one obtains
\bea
\label{eq:quartic:App:interm3}
\frac{\dd \alpha}{\dd N} &=  \frac{\Gamma \left( \frac{1}{4} \right)}{2\Gamma \left( \frac{3}{4} \right)} \left( \frac{\lambda}{H^2}\alpha^{1/2} - \frac{3H^2}{8\pi^2} \alpha^{3/2}\right) \,.
\eea
Notice that if one had directly integrated \Eq{eq:quartic:App:interm2} over $\sigma$ and substituted \Eq{eq:moments}, one would have obtained a trivial relationship, which is why we first multiplied \Eq{eq:quartic:App:interm2} by $\sigma^2$ before integrating over $\sigma$.

If the inflaton potential is of the plateau type and  $H$ can be approximated by a constant, this equation can be solved and one finds
\bea
\label{eq:solution:alpha}
\alpha\left(N\right)=\frac{8\pi^2\lambda}{3H^4}\tanh^2\left\lbrace \sqrt{\frac{3\lambda}{2}}\dfrac{\Gamma\left(\frac{1}{4}\right)}{8\pi\Gamma\left(\frac{3}{4}\right)}\left(N-N_0\right)+\mathrm{arctanh}\left[\sqrt{\frac{3H^4\alpha\left(N_0\right)}{8\pi^2\lambda}}\right] \right\rbrace\, ,
\eea
which gives rise to \Eq{eq:quartic:plateau:Appr} for the second moment $\langle \sigma^2 \rangle$.

If the inflaton potential is monomial, the function $H(N)$ is given by \Eq{eq:Hubble} and although an analytical solution still exists, it is less straightforward to derive. The first step consists of writing \Eq{eq:quartic:App:interm3} in terms of an equation for $\langle\sigma^2\rangle$ using \Eq{eq:moments},
\bea
\frac{\dd \langle \sigma^2 \rangle}{\dd N} &= -\frac{2}{3}\left[\frac{\Gamma \left( \frac{1}{4} \right)}{\Gamma \left( \frac{3}{4} \right)}\right]^2\frac{\lambda \langle \sigma^2\rangle^2}{H^2} + \frac{H^2}{4\pi^2}  \, .
\eea
The next step is to use $x\equiv H/H_\uend$ as a time variable, which gives rise to
\bea
\label{eq:Ricatti}
\frac{\dd \langle \sigma^2 \rangle}{\dd x} = \frac{2}{3}\left[\frac{\Gamma \left( \frac{1}{4} \right)}{\Gamma \left( \frac{3}{4} \right)}\right]^2\frac{\lambda x^{4/p-3}}{H_\uend^2}\langle \sigma^2\rangle^2 - \frac{H_\uend^2x^{4/p+1}}{4\pi^2}  \,.
\eea
This equation is of the Ricatti type and can be transformed into a second-order linear differential equation making use of the change of variables
\bea
\label{eq:gtrans}
\langle \sigma^2 \rangle= -\frac{3}{2}\frac{H_\uend^2}{\lambda}\left[ \frac{\Gamma \left( \frac{3}{4} \right)}{\Gamma \left( \frac{1}{4} \right)}\right]^2 x^{3-\frac{4}{p}} \frac{1}{f(x)}\frac{\dd f}{\dd x}\, .
\eea
By plugging \Eq{eq:gtrans} into \Eq{eq:Ricatti}, one obtains
\bea
\frac{\dd^2 f}{\dd x^2} + \left( 3- \frac{4}{p}\right)\frac{1}{x} \frac{\dd f}{\dd x} - \frac{\lambda}{6\pi^2}\left[ \frac{\Gamma \left( \frac{1}{4} \right)}{\Gamma \left( \frac{3}{4} \right)}\right]^2x^{8/p-2}f = 0\,.
\eea
This equation can be solved in terms of modified Bessel functions of the first kind $I$. Making use of \Eq{eq:gtrans}, the solution one obtains gives rise to
\bea
\label{eq:quartic:App:interm10}
\langle \sigma^2 (x)\rangle =& \frac{x^{2-4/p}}{2A}\left( 2 - \frac{4}{p} \right)
\\ &
- \sqrt{\frac{B}{A}}\frac{x^2}{2}\left\lbrace \frac{  I_{-\frac{p}{4}-\frac{1}{2}}(W) + I_{-\frac{p}{4}+\frac{3}{2}}(W)  + C \left[ I_{\frac{p}{4}+\frac{1}{2}}(W) + I_{\frac{p}{4}-\frac{3}{2}}(W) \right]}{I_{\frac{p}{4}-\frac{1}{2}}(W) + C I_{-\frac{p}{4}+\frac{1}{2}}(W)} \right\rbrace \,,
\eea
where we have defined
\bea
A = \frac{2}{3}\left[ \frac{\Gamma \left( \frac{1}{4} \right)}{\Gamma \left( \frac{3}{4} \right)}\right]^2 \frac{\lambda}{H_\uend^2}\,, \qquad \ B = \frac{H_\uend^2}{4\pi^2}\,, \qquad \ W = \frac{p}{4}\sqrt{AB} x^{4/p}\, ,
\eea
where $C$ is an integration constant that can be set as follows: In the asymptotic past, $W\gg 1$ and the Bessel functions can be expanded in this limit, $I_\alpha (W) \simeq \ee^W/\sqrt{2\pi W}$. Unless $C=-1$, the term inside square brackets in \Eq{eq:quartic:App:interm10} goes to $1$ and one finds $\langle\sigma^2\rangle\simeq -\sqrt{B/A}x^2/2 <0$ which would not be consistent. As a consequence, $C=-1$ is the only choice that allows the solution~(\ref{eq:quartic:App:interm10}) to be defined over the entire inflationary period. Setting $C=-1$, \Eq{eq:quartic:App:interm10} can be simplified and one obtains
\bea
\langle \sigma^2 (N)\rangle =\frac{\Gamma\left(\frac{3}{4}\right)}{\Gamma\left(\frac{1}{4}\right)}\sqrt{\frac{3}{2\lambda}} \frac{H^2}{2\pi} \frac{K_{\frac{p}{4}+\frac{1}{2}}(W)}{K_{\frac{p}{4}-\frac{1}{2}}(W) } \,,
\eea
where $K$ is the modified Bessel function of the second kind.
\bibliographystyle{JHEP}
\bibliography{StochasticSpectator}
\end{document}